\pdfoutput=1
\documentclass[final]{cvpr}

\usepackage{times}
\usepackage{epsfig}
\usepackage{graphicx}
\usepackage{amsmath}
\usepackage{amssymb}

\usepackage{xcolor}
\usepackage{xspace}
\usepackage{boldline} 
\usepackage[inline]{enumitem}
\usepackage[normalem]{ulem}
\usepackage{booktabs}
\usepackage{graphicx}

\def\cvprPaperID{****} 
\def\confYear{CVPR 2022}

\begin{document}

\title{ISP-Agnostic Image Reconstruction for Under-Display Cameras}

\author{Miao Qi \qquad Yuqi Li \qquad Wolfgang Heidrich \\ 
KAUST Visual Computing Center \\
Saudi Arabia\\
{\tt\small miao.qi@kaust.edu.sa \qquad liyuqi1@nbu.edu.cn \qquad wolfgang.heidrich@kaust.edu.sa}
}


\maketitle

\begin{abstract}
   Under-display cameras have been proposed in recent years as a way to
  reduce the form factor of mobile devices while maximizing the screen
  area. Unfortunately, placing the camera behind the screen results in
  significant image distortions, including loss of contrast, blur, noise,
  color shift, scattering artifacts, and reduced light sensitivity.

  In this paper, we propose an image-restoration pipeline that is
  ISP-agnostic, i.e. it can be combined with any legacy ISP to produce
  a final image that matches the appearance of regular cameras using
  the same ISP. This is achieved with a deep learning approach that
  performs a RAW-to-RAW image restoration.

  To obtain large quantities of real under-display camera training
  data with sufficient contrast and scene diversity, we furthermore
  develop a data capture method utilizing an HDR monitor, as well as a
  data augmentation method to generate suitable HDR content. The
  monitor data is supplemented with real-world data that has less
  scene diversity but allows us to achieve fine detail recovery
  without being limited by the monitor resolution. Together, this
  approach successfully restores color and contrast as well as image
  detail.
\end{abstract}

\newcommand{\HW}{Huawei\xspace}
\newcommand{\todo}[1]	{{\textcolor{red}{Todo: #1}}}
\newcommand{\note}[1]   {{\textcolor{orange}{Note: #1}}}
\newcommand{\Miao}[1]    {{\textcolor{blue}{Miao: #1}}}
\newcommand{\remove}[1] {}  
\newcommand{\beginsupplement}{%
        \setcounter{table}{0}
        \renewcommand{\thetable}{S\arabic{table}}%
        \setcounter{figure}{0}
        \renewcommand{\thefigure}{S\arabic{figure}}%
        \setcounter{section}{0}
        \renewcommand{\thesection}{S\arabic{section}}%
     }

\section{Introduction}


\begin{figure*}[h]
  \centering
    \includegraphics[width=\textwidth]{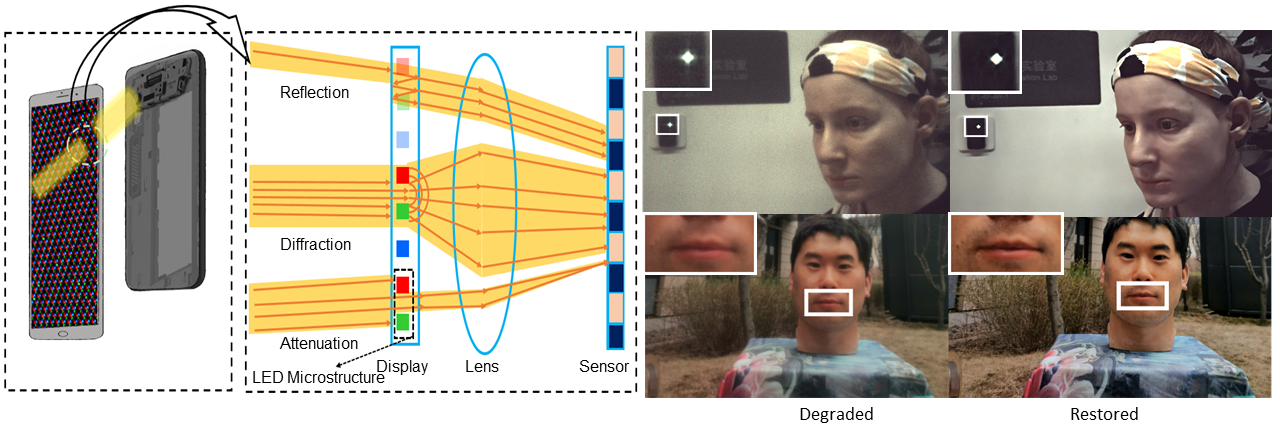}
  \caption{Under-display camera (UDC) optics and example images. The
    left figure is an illustration of optical distortions created by
    under display cameras. Three main optical effects cause image
    degradation in UDC cameras: a) blocking and absorption of light in
    the display results in brightness and color distortions. b) the
    regular pixel structures of the display act as a diffraction
    grating that haze-like low frequency blur as well as stripe
    artifacts around light sources. c) Multiple reflection between the
    display layers results in a relatively compact blur.  The right
    figure shows the resulting image degradation as well as our
    restoration result.  We can see all of the mentioned degradations
    in the right figure, especially the ``stripe-like'' artifact of
    the green spot light in the right top image. We removed it very
    successfully in the restored image. In the right bottom, degraded
    image contrast become low and the man's mustache detail is blurred
    and lost in the degraded image, but all restored in our restored
    image.}  
  \label{fig:teaser}
\end{figure*}

Ever since the invention of smart phones, there has been a push to
maximize the available screen area as a percentage of the total device
surface area. As a result, most other components have now been removed
completely from the front face of the phone, with one notable
exception: the user-facing, or ``selfie'' camera, which is usually
placed in a ``notch'' or hole cut out from the display.

For aesthetic and usability reasons, it has been proposed that the
selfie camera could be moved under the display~\cite{park200842}.
This concept would make use of the fact that OLED displays can be
embedded in a transparent substrate. Since the actively emitting area
of an OLED pixel is much smaller than the pixel spacing, sufficient
light may pass through the display to reach the camera.

Unfortunately, the commercial adoption of this concept has been
relatively slow due to the degradation in image quality (see
Fig.~\ref{fig:teaser}). Specifically, the regular pixel structures in
the display act as a diffraction grating that distorts the captured
image~\cite{toisoul2017practical}. We can observe some amount of blur,
a ``haze''-like reduction in contrast (due to a blur kernel with a
long tail), as well as significant color distortion from the
absorption in the display.

Image reconstruction for this type of under-display camera (UDC)
therefore entails deblurring, dehazing, color restoration, as well as
handling increased noise due to the reduced amount of light incident
on the camera. Although the distortion appears to be largely
shift-invariant, this remains a difficult reconstruction problem.

For practical deployment, there is another major consideration: smart
phone manufacturers have invested a significant amount of effort into
developing image signal processing (ISP) pipelines for their
products. These are complicated systems that achieve a certain
brand-specific look, which we would like to preserve. One way to
achieve this goal would be to attempt to train a deep network to not
only perform the image reconstruction, but also emulate a given ISP at
the same time. However, this approach requires that the deep network
not only learns the UDC reconstruction task but also an immensely
complex black-box ISP system, which would require infeasible amounts
of training data (images captured with an UDC and a comparison regular
camera). Instead, we opt for an ISP-agnostic design enabled by a
raw-to-raw training pipeline. Please see
Section~\ref{sec:agnostic-design} for a more detailed discussion of
this design choice.

Our RAW-to-RAW reconstruction framework is based on a Wasserstein
generative adversarial network with gradient penalty (WGAN-GP) deep
network architecture~\cite{gulrajani2017improved}, which can be
successfully trained to deal with the distortions of the UDC images.
We also devise a new HDR monitor-based capture setup that allows us to
easily capture image pairs with a UDC camera and a regular reference
camera. This setup allows us to train the network on the overall
reconstruction task with a large scene diversity with challenging high
contrast scenes that would be exceedingly difficult and cumbersome to
capture ``in the wild''. On the other hand, any monitor-based capture
system is inherently limited by the pixel count of the display; we
therefore supplement this data with a limited real-world dataset that
has much reduced scene diversity but full resolution. 
In summary, we make the  following technical contributions:

\begin{itemize}
\item In order to preserve the brand-specific ISP feature, we develop
  an ISP-agnostic RAW-to-RAW pipeline. Ours is the first work to
  tackle the UDC reconstruction problem in a RAW-to-RAW fashion.
 
\item We propose a HDR data augmentation method for HDR portrait
  images.
  
\item We propose a series of data collection methods and corresponding
  preprocessing methods to create a large dataset of high-quality
  under-display camera images. 

\item We test our algorithm on both monitor data and real-world data.

\end{itemize}

\section{Related Work}
\label{sec:related}

Image restoration from under-display cameras is a fairly recent
problem, which is only considered by a small number of peer reviewed
publications so far~\cite{zhang2020image}. However, the problem bears
similarity to other image reconstruction tasks that have been more
thoroughly researched, including dehazing, deblurring, denoising, and
color correction.

\paragraph*{Image Dehazing and Deblurring.}
An early example of image de-hazing was based on manually designed
image priors, such as the highly successful dark channel
prior~\cite{he2010single}. More recently, the attention has shifted to
deep learning approaches. DehazeNet~\cite{cai2016dehazenet} was
directly inspired by the dark channel prior in their choice of the
``Maxout'' activation function. Li et al.~\cite{li2017aod} proposed a
CNN based on the atmospheric scattering model and designed an
end-to-end neural network. Chen~\cite{chen2019gated} introduced
dehazing CNN with a ResNet structure, featuring gated fusion with a
multiscale approach, while Guo~\cite{guo2019dense123} utilized a
dense-connection encoder-decoder network operating on each color
channel independently.

Multiscale processing is very useful also in image deblurring. Nah et
al.~\cite{nah2017deep} proposed a multiscale deblur neural
network structure and loss function. Kupyn et
al.~\cite{kupyn2018deblurgan} introduced the DeblurGAN, which can
achieve a state-of-art deblurring result.
  
\paragraph*{Imaging Through Scattering Media.}
A closely related problem is that of imaging through scattering media,
which is usually tackled with a more model-based approach, for example
by measuring the transmission matrix~\cite{popoff2010measuring}. The
scattering inside optically thin media exhibits a ``memory
effect''~\cite{feng1988correlations}, and within the region of this
memory effect the blur kernel is shift-invariant, so that the image
formation model simplifies to a convolution. At the same time, the
blur kernels exhibit long tails, creating a haze-like reduction of
contrast~\cite{katz2014non}. These results are directly applicable to
the restoration problem for under-display cameras, since the display
layer acts as a weakly scattering diffuser. This is similar to the
contrast reduction observed in recent works on diffractive optical
elements~\cite{peng2019learned}.

\paragraph*{Deep Learned ISPs.}
Recently, Ignatov~\cite{ignatov2020replacing} proposed to train a
single deep network to replace legacy ISPs. While their results are
promising, they also highlight the difficulties in trying to fit such
large, complex software systems with single network: the authors
report PSNR values only around 21~dB. The most recent results from the
neural network ISP challenge~\cite{ignatov2021learned}, report PSNR
values of up to 24~dB, which is still not considered sufficient for
commercial deployment. This observation is one reason why we decide on
an ISP-agnostic design (Sec.~\ref{sec:agnostic-design}.)

\paragraph*{ISP-Dependent and Monitor Data UDC Restoration}
At ECCV 2020, Zhou et al. for the first time held a UDC restoration
challenge, resulting in four
contributions~\cite{zhou2020udc,yang2020residual,sethumadhavan2020transform,sundar2020deep}.
The dataset for the challenge is captured on machine vision camera
with a simulated UDC hardware.  Because of the use of a machine vision
camera, the dataset does not involve complicated ISP process as it
would on smart phone systems, which is where real UDC hardware would
likely be deployed. Instead, our dataset is based on a real smartphone
camera with a UDC configuration.

Furthermore, the data from the challenge is captured exclusively on a
standard dynamic range (SDR) monitor. This is problemetic since many
of the artifacts caused by diffraction are actually only visible in
high contrast scenes and around light sources (see
e.g.~\cite{rouf2011glare,sun2020learning}). Such imagery cannot be
created on an SDR monitor. Furthermore, all monitor data is limited by
the pixel resolution of the screen, an we demonstrate in this work
that this limits the overall achievable reconstruction quality on real
world data. In our work, we resolve these issue by combining HDR
monitor data with a real-world dataset that can be used to fine-tune
the recovery of small high frequency features.

Most recently, Yang et al.~\cite{yang2021designing} proposed a screen
design optimization for the UDC screen which improves the UDC image
quality by choosing a randomized layout for the LED subpixel
structures in the display. This work is orthogonal to our approach,
and the two methods could easily be combined.

\paragraph*{RAW-to-RAW Pipelines}
Recently, some RAW-to-RAW pipelines have been introduced for image
restoration tasks such as super
resolution~\cite{xu2019towards,zhang2019zoom} and denoising
~\cite{abdelhamed2018high,cao2021pseudo}. However, to the best of our
knowledge, we are the first to tackle the UDC restoration problem in
this fashion.


  \begin{figure*}[t!]
  \centering
  \includegraphics[width=\linewidth]{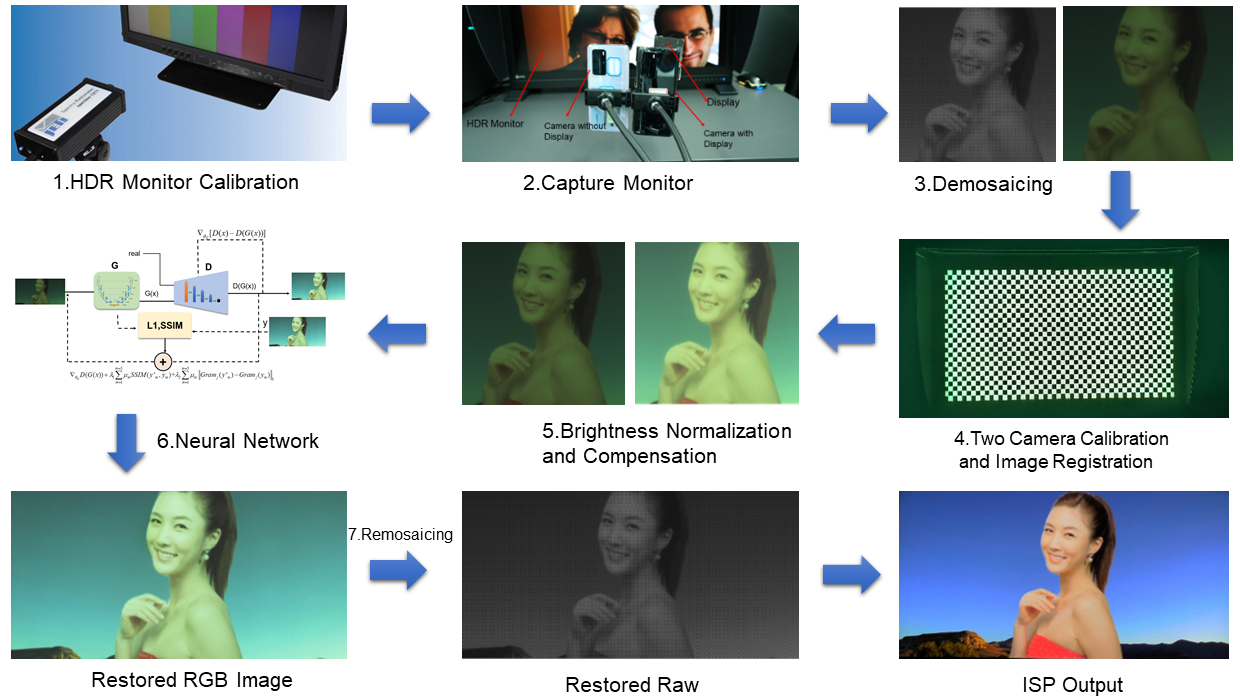}
  \caption{Our RAW-to-RAW pipeline. Our pipeline includes the
    following steps: 1. calibration of the HDR monitor to produce
    colors and intensities matching real-world scenes; 2. capture of
    image pairs with two smartphones (one regular camera, and one UDC
    camera); 3. simple interpolation-based demosaicking to get three
    color channels for each pixel; 4. image registration by
    homography; 5. brightness normalization; 6. WGAN-GP based image
    reconstruction; and 7. re-apply a Bayer filter pattern to convert
    the image back to raw format. After this pipeline, a legacy ISP
    can be used to obtain the final image.}
  \label{pipeline}
\end{figure*}

\section{Problem Description and System Design}

\subsection{UDC Image Formation Model}

Placing a transparent display on top of a camera degrades the image in
a number of ways (compare Fig.~\ref{fig:teaser}). First, the display
absorbs some of the light, and this absorption is
wavelength-dependent. Therefore we expect both lower light sensitivity
(i.e. increased noise) and color distortions. Second, the pixels of
the display form a regular grid of micro-structures that act as a
diffraction grating, which amounts to a chromatic blur that is mostly
shift invariant except in the case of lens
distortion~\cite{rouf2011glare,yang2021designing}. Finally, the front
and back surface of the display can create multiply reflected light
paths, which also result in image blurring with a shift-invariant
PSF. The total model can be expressed as
\begin{equation}
  i_k(x,y)=\int c_k(\lambda)d(\lambda) [s(x,y,\lambda)*o(x,y,\lambda)]d\lambda+n(x,y)
\end{equation}
where $o_k(x,y)$ is channel $k$ of the intrinsic object/scene that
needs to be reconstructed
\begin{equation}
  o_k(x,y)=\int c_k(\lambda)o(x,y,\lambda)d\lambda,
\end{equation}
while $s_k(x,y,\lambda)$ is the
corresponding full system point spread function(PSF). $c_k(\lambda)$
is the absorption coefficient of color channel $k$ in the sensor for a
given wavelength, and $d(\lambda)$ is the absorption of the OLED
display. $n(x,y)$ is the system noise, while $i_k(x,y)$ is the image
captured by the camera sensor. Reconstructing $o_k(x,y)$ from
$i_k(x,y)$ is an ill-posed inverse problem.

Especially the diffractive part of the blur causes issues, since it
can have a long ``tail'' due to higher diffraction
orders~\cite{sun2020learning}, and it is also strongly
wavelength-dependent, while the sensor data has only three color
channels instead of full spectral information. This makes it necessary
to employ deep learning image reconstruction approaches instead of
older optimization-based approaches.



\subsection{ISP-Agnostic Pipeline}
\label{sec:agnostic-design}

There three main options for how image reconstruction tasks can
interact with existing ISP software: first we could design an
RGB-to-RGB pipeline, that takes as input an ISP processed image and
outputs another RGB image that hopefully mimics the appearance of a
regular, ISP-processed camera image without UDC hardware. This
approach is complicated by the complexity of modern ISPs that perform
scene-dependent processing (e.g. for portraits vs. landscapes, dark
vs. bright environments etc.), as well as non-linear processing such as
lookup tables. Because UDC image and reference training images could be
processed differently by the ISP, training a network to be able to work
with all different combinations of internal choices in the blackbox
ISP would require prohibitive amounts of training data.

A second design choice would be a RAW-to-RGB pipeline, which takes as
input a raw image and produces an output that looks like an ISP
processed image of a reference camera. While slightly easier than the
first design choice, this approach still conflates the learning  of
the UDC restoration problem with the blackbox learning of the ISP
module, and as discussed in Section~\ref{sec:related}, even just the
latter task is difficult enough that there is currently no truly
satisfactory solution.

The final design choice is to employ a RAW-to-RAW pipeline, which
completely sidesteps the ISP complexities, and allows the neural
network to focus exclusively on the UDC reconstruction task. After raw
image restoration, any existing legacy ISP can be applied to achieve
the desired look in the final image.  Figure~\ref{pipeline} shows an
overview of our pipeline.

\section{Network Architecture and Loss}
\label{sec:network}


As shown in the supplementary Fig.~\ref{fig:network}, for the neural network we
choose the WGAN-GP structure~\cite{gulrajani2017improved}. This choice
is motivated by a number of considerations. First, GANs excel at
preserving and recovering texture detail~\cite{kupyn2018deblurgan},
which is required in UDC cameras due to the display-induced blur. GANs
have also shown to work very well on style transfer and other image to
image mappings~\cite{zhu2017unpaired}, which is a task similar to the
color and contrast restoration problem in UDC images. As a
semi-supervised learning method, GANs also generally minimize the
amount of training data needed, which is advantageous in our
setting. Finally, the WGAN-GP in particular is very stable and easy to
train~\cite{gulrajani2017improved}.

The WGAN-GP loss can be written as:
   \begin{equation}\label{GAN loss}
  		\mathcal{L}_a=\mathop{\mathbb{E}}_{x \sim \mathbb{P}_r}[D(x)]-\mathop{\mathbb{E}}_{x' \sim \mathbb{P}_g}[D(x')]+\lambda_g \mathop{\mathbb{E}}_{x' \sim \mathbb{P}_{\hat{x}}}[(\Vert \nabla_x'D(x') \Vert_2 -1)],
  \end{equation}
where $\mathbb{P}_r$ is the real image distribution, $ \mathbb{P}_g$
is the generated image distribution, and $D(x)$ represents the output
of the discriminator. The last term is the gradient penalty term.

Since our training images have slight variations in the brightness
mapping, we use both the SSIM loss and the perceptual loss to compare
images, since SSIM is less sensitive to such
differences~\cite{zhao2016loss}. The combined content loss is then:
 \begin{equation}
  \label{content loss}
  \mathcal{L}_c=\lambda_1 SSIM(y'-y) +\lambda_2 \Vert Gram_j(y')-Gram_j(y) \Vert_1,
 \end{equation}
where $y$ is the ground truth degraded image. $y'$ is the restored image.
$\lambda_1$ and $\lambda_2$ are the weights for the SSIM and
perceptual loss. $Gram_j$ is the Gram matrix defined by:
\begin{equation}
  Gram_j(y)=F_j(y)F_j(y)^{T}/C_jH_jW_j,
\end{equation}
and $F_j(x)$ is the VGG19 $j$-th layer output feature. $C$, $H$, $W$
are of the size of the features. Here we choose the $j=15$.
In the end, the overall loss is:
\begin{equation} \label{all loss}   
  		\mathcal{L}_{total}=\mathcal{L}_a+ \mathcal{L}_c,
\end{equation}

\paragraph*{Generator}

\begin{figure}[h]
  \centering
  \includegraphics[width=\linewidth]{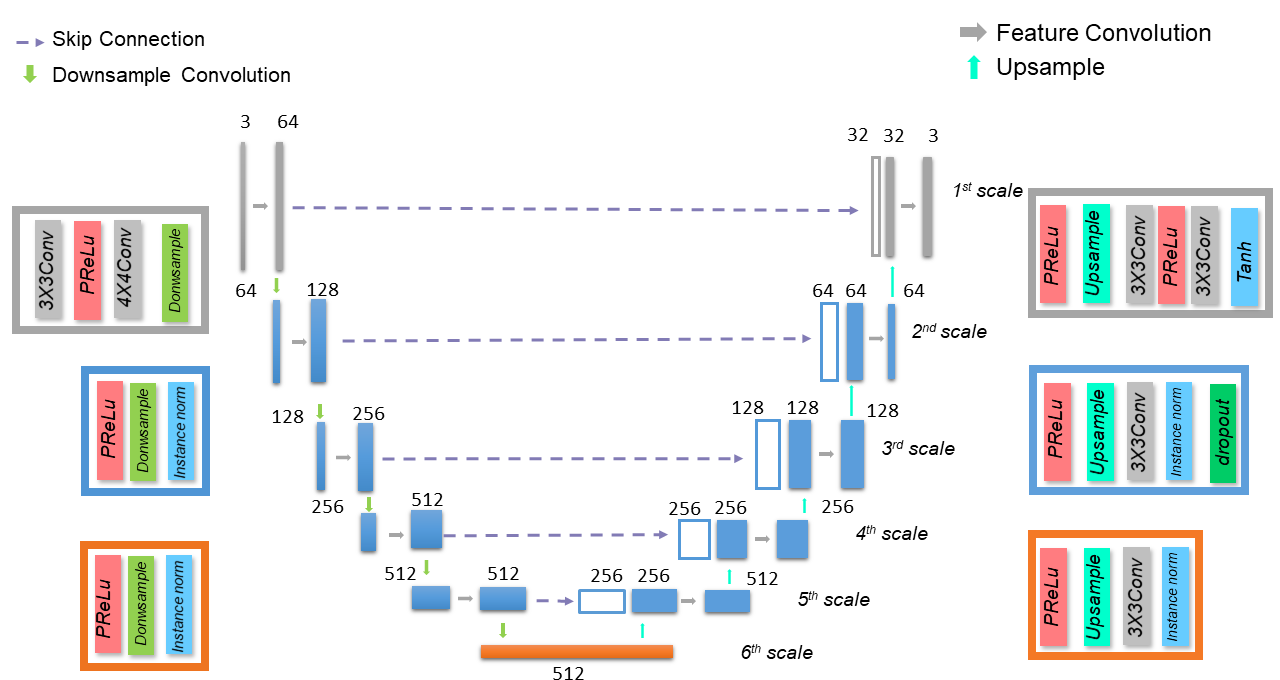}
  \caption{The generator network. The architecture was carfully tuned
    to the UDC problem (see text). The color outlines indicate the
    structure of correspondingly colored blocks in the encoder and
    decoder. Because the pooling layer will lose some information, we
    replace the pooling layer with convolution with stride of 2.}
  \label{UNet}
\end{figure}

The generator is an encoder-deconder structure as shown in Fig.~\ref{UNet}. The following are the tailored improvements for our task to the generator:
\begin{enumerate*} [label=\itshape\alph*\upshape)]
\item We replace the pooling layer with stride 2 downsample
  convolution layer. Because in our task, the PSF is shift-invariant,
  our convolution kernel does not need a large receptive field, and the
  pooling layer will cause some information loss. By using this
  downsample convolution layer, we can not only shrink the image size
  and match it with different scales, but also counteract the loss of
  information.
\item We replace the transpose convolution with upsampling. This can
  avoid block artifacts (see supplement Fig.~\ref{checkernoise}).
\item In the encoder part, we use a bigger convolution kernel
  ($4\times4$) to have a broader view. In the decoder part, we use the
  smaller convolution kernel ($3\times3$) to preserve the fine detail.
\item Because our training batch is small, the variance of different
  batches is large. Therefore, we use the instance norm.

\end{enumerate*}

\paragraph*{Discriminator} We use a simple CNN as a discriminator.
Since the WGAN-GP architecture does not allow for the use of the batch
norm, we e use the instance norm instead. For the activation function,
we use the LeakyReLu as suggested by Gulrajani et
al.~\cite{gulrajani2017improved}. The convolution kernel is
$4\times 4$.

\section{Training Dataset}
\label{sec:Training Dataset}

For training the network we require image pairs composed of one image
from a UDC camera and one image from a reference camera that does not
have a screen in front of it, but is otherwise identical. 
As mentioned in the introduction, our training data is composed of two
types of data: image pairs captured of an HDR monitor, and real world
data.

\subsection{Monitor Dataset}
\label{sec:monitor}
\subsubsection{Data Collection Setup}

The purpose of the monitor data is to be able to cover a large
diversity of scenes such as both indoor and outdoor scenes,
including those specific to a range of geographic
locations. Traveling to such a variety of environments with a UDC and
reference camera would be impractical. For quality training we also
require excellent pixel-to-pixel alignment between the UDC and
reference image, which is easily achieved with a homography on monitor
data, but can be difficult for real world scenes, especially those
with high depth complexity (also see Section~\ref{sec:realworld}).

The monitor capture setup is centered around an Eizo CG3145 HDR
monitor with a resolution of $4,096 \times 2,160$. This monitor uses
the dual modulation principle~\cite{seetzen2004high} to generate a
typical contrast ratio of 1,000,000:1. However, unlike most most dual
modulation HDR monitors that use an LCD illuminated by a low-frequency
LED backlight~\cite{seetzen2004high}, the Eizo CG3145 actually uses
two LCD layers stacked ontop of each other on the same glass
substrate. This allows the monitor to achieve not only high global
contrast, but also excellent local contrast of high frequency
features. This makes the monitor ideally suited for re-photography and
similar tasks. Although the monitor can not achieve a {\em peak brightness}
comparable to sunny outdoor scenes, it can achieve the peak {\em
  contrast} of the vast majority of real world environments, and as
such can realistically generate glare effects due to the diffraction
in the UDC setup.

\subsubsection{HDR Portrait Dataset}

Since UDC cameras are designed to be facing the user, selfies and
group selfies and group selfies (``groufies'') are the primary type of
images the camera is intended for.  We took this data bias into
account bias and generated a large amount of portrait images for
display. Selfies are often taken under challenging lighting
situations, with people posing in front of bright, sunlit landscapes
or in dark restaurants. Our use of an HDR monitor is motivated
by this observation, as well as by the realization that diffraction
artifacts often show only in high contrast scenes.

Unfortunately, there is no publicly available database of HDR portrait
or selfie photos with different real world backdrops. We therefore
decided to create our own dataset by compositing SDR portraits into
HDR background images (see the Supplement
Fig. \ref{fig:composite}). To this end, we collected five thousand
high-quality human half-length portraits with masks from the public
matting human
dataset\footnote{\url{https://github.com/aisegmentcn/matting\_human\_datasets}}. We also
collect two thousand HDR images from the HDRIHAVEN dataset.  We
treated the portrait and HDR images as foreground and background,
respectively, and fused them to generate display images using
alpha-matting.  Finally, we manually selected 850 credibly looking
composites to serve as content to be shown on the HDR monitor (764 for
training and 86 for testing).  More details as well as example images
of the dataset can be found in the Supplement.

Note that a straightforward compositing of the to datasets mixes
different lighting conditions for the foreground portraits and the
background environments. However, this is not a concern in our
setting, since such situations can also occur in real images,
e.g. when photographing from a shaded region into a sunlit background.
Furthermore in our RAW-to-RAW pipeline any potential scene dependent
decisions of an ISP are absent so that the restoration of UDC images
should be largely independent of specific scene composition. Finally,
the manual screening eliminates the most unnatural looking composites.

\subsection{Real World Dataset}
\label{sec:realworld}
While the monitor data provides a large diversity of scene content and
illumination scenarios, its spatial resolution is limited by the 2
Mpixel resolution of the HDR display. To finetune the network training
for full camera resolution, we therefore also captured a real world
dataset with a much more limited diversity of scenes. 

These image pairs were captured on a tripod-mounted camera with and
without the UDC hardware. Overall we collected 1020 image pairs,
mostly indoors. We performed automatic pixel-to-pixel alignment on all
image pairs, by first extracting the image features of UDC and
reference images. Based on these features, we computed the homography
matrix~\cite{malis2007deeper} and applied it to the UDC image. In
addition, we computed the optical flow between the two images and
warped the reference image accordingly for a pixel-to-pixel alignment.

Although this process produces good results for scenes of moderate
complexity, it does not work sufficiently well on all images. We
therefore manually selected 620 images where the alignment process was
considered successful (520 for fine-tuning the training and 100 for
testing).

Note that automatic pixel-to-pixel image alignment tends to fail in
difficult lighting situations or when the depth complexity of the
scene is too high. Overall this results in a limited scene diversity
in the real world dataset (indoor scenes with good illumination and
not too complex object geometry). As a result, this dataset is
suitable for fine-tuning but not for training the UDC reconstruction
process from scratch.


\subsection{Data Augmentation}\label{Data Augmentation}
We performed data augmentation for the training data. In addition to
the standard crop, flip, transpose
operations~\cite{iandola2016squeezenet}, we also introduced resizing
and brightness augmentation. Resizing augmentation was done by
resizing the degraded and ground truth image to different scale.

The brightness augmentation was done by multiplying a factor number to
the pixel value of degraded and ground truth image. For details,
please refer
to the Supplement.

\subsection{Training and Test Detail}
As mentioned above, we use the monitor dataset for pre-training, to
obtain a large scene diversity, and then fine-tune using the real
world data.  Because of the limited GPU memory, we divided the input
image pairs into $512 \times 512$ tiles. The resizing augmentation was
used to achieve multiscale training. We normalized the input data to
$[-1,1]$.  The Adam optimizer was used with learning rate 0.9 and
weight decay 0.999.  The batch size is 2. We use a lower weight for the perceptual loss and a higher one for
SSIM, which is less sensitive to brightness differences. We find that $\lambda_g=10$
in Eq. \ref{GAN loss} and SSIM weight $\lambda_1=50$ and perceptual loss weight $\lambda_2=30$ in
Eq.\ref{content loss} gave the best performance.

\section{Experimental Results}

\subsection{Ablation Study}

\begin{table}[]
\centering
\resizebox{0.47\textwidth}{!}{%
\begin{tabular}{@{}ccccccccc@{}}
\toprule
\multicolumn{2}{c}{Data Augmentation}                                          &  & \multicolumn{3}{c}{Loss Function}         &  & PSNR  & SSIM \\ \midrule
\begin{tabular}[c]{@{}c@{}}Brightness\\ Augmentation\end{tabular} & Resize     &  & SSIM       & L1         & Perceptual Loss &  &       &      \\ \midrule
\checkmark                                                        &            &  & \checkmark & \checkmark &                 &  & 32.23 & 0.76 \\
\checkmark                                                        & \checkmark &  & \checkmark & \checkmark &                 &  & 33.76 & 0.88 \\
                                                                  &            &  & \checkmark &            & \checkmark      &  & 34.09 & 0.85 \\
\checkmark                                                        &            &  & \checkmark &            & \checkmark      &  & 34.22 & 0.88 \\
\checkmark                                                        & \checkmark &  & \checkmark &            & \checkmark      &  & {\bf 35.72} & {\bf 0.92} \\ \bottomrule
\end{tabular}%
}
\caption{Data augmentation and loss function ablation study}
\label{Ablation Study}
\end{table}

To understand the impact of our data augmentation and loss function
contribution, we conduct an ablation study. We incrementally added one
improvement to each case at a time.  As shown in Table~\ref{Ablation
  Study}, the biggest gain comes from the resizing data
augmentation. As shown in the first and second row, there is a gain of about 1.5 dB in PSNR and about 0.12
improvement in the SSIM. The reason is that we train on cropped images
in order to save memory and time. However, this may make it harder for
the network to learn global structures. The resizing augmentation
counteracts this problem by establishing a multi-scale strategy.

The brightness augmentation is necessary because the final test image
is captured in auto exposure mode and with all kind of different
scenes. The brightnesses vary over a wide range. In order to test the
model performance, we also include some images captured in the extreme
exposure settings, including both under- and over-exposed
scenes. Without such brightness augmentation, the learning tends to
result in over-fitting. See our ablation study (Table~\ref{Ablation
  Study}), rows 3 and 4. The SSIM is improved by about 0.03.

Unlike most conditional GAN papers~\cite{isola2017image} that choose
the L1 loss, we choose the perceptual loss as part of the content
loss. The rationale and more detail can be found in the Supplement.

In order to illustrate why we choose the RAW-to-RAW pipeline, we
performed an experiment to let the neural network learn in the
ISP-processed image dataset directly. The result is in the Supplement.
  As shown in Supplement Fig.~\ref{Direct Training Result of JPG Image}, the output result is even worse than
the input. This is because in our case, the difference between the
degraded image and ground truth image caused by UDC is even less than
the difference caused by unpaired brightness and color. The tendency
is that the network tries to learn brightness and color features
first. However, this change is so evident that it will also mislead
the neural network to ``learn'' the wrong features. The brightness and
color changes are so unpredictable that usually the neural network
ends up not affecting the final image at all.

After the restoration, we applied a Bayer pattern to the output RGB
image. This converts the restored RGB image to a raw image format
again, which can then be processed by the standard cell phone ISP (or
any other desired ISP). In the following we show our pipeline
inference result before ISP and after ISP in different datasets.

\begin{table}[]

\centering
\resizebox{0.47\textwidth}{!}{%
\begin{tabular}{@{}lccccccccc@{}}
\toprule
               & \multicolumn{4}{c}{Raw Image}    &  & \multicolumn{4}{c}{ISP-proceed Image} \\ \cmidrule(lr){2-5} \cmidrule(l){7-10} 
               & PSNR  & SSIM & LPIPS & CIEDE2000 &  & PSNR   & SSIM  & LPIPS  & CIEDE2000 \\ \midrule
Degraded Image & 22.85 & 0.84 & 0.336       & 0.08          &  & 22.85  & 0.85  & 0.154       & 6.32         \\
Restored Image & 33.95 & 0.95 &   0.133    &  0.02         &  & 34.81  & 0.94  &  0.048      & 1.83          \\ \bottomrule
\end{tabular}%
}
\caption{Monitor dataset restoration results. The CIEDE2000~\cite{luo2001development} is the average value.}
\label{The Monitor Dataset Restoration Result}
\end{table}

\begin{figure*}[h]
  \centering  
  \includegraphics[width=\linewidth]{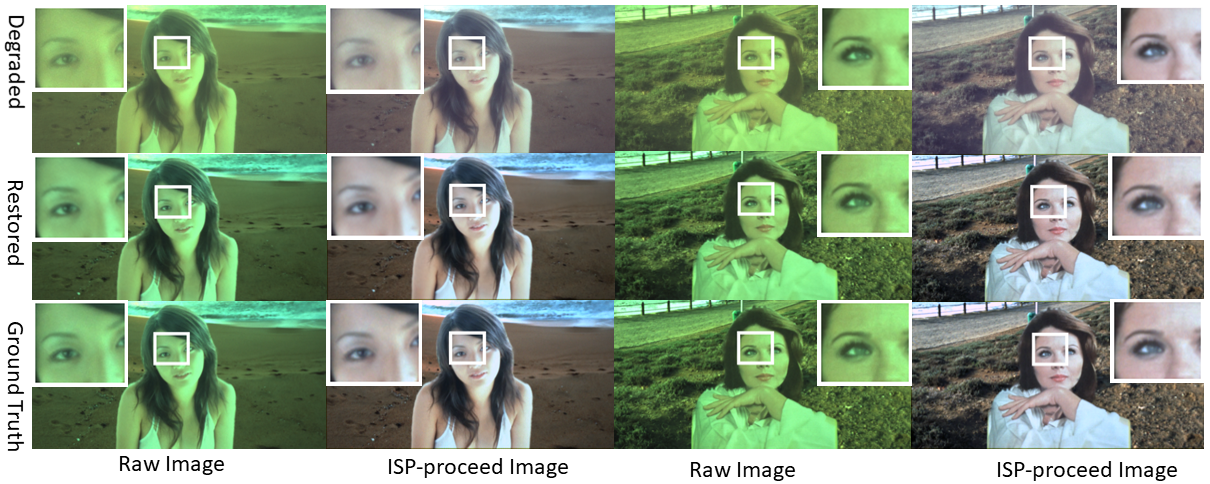}
  \caption{Monitor raw data and ISP-processed data results. The green
    image is the demosaicked raw image. The ISP-proceed image is the
    restored raw image that after passing through the ISP. Our
    restored UDC image matches with the ISP-processed image of the
    reference (non-UDC) camera. }
  \label{Monitor Raw Data and After ISP Data Result}
\end{figure*}

The restored, ISP-processed image for monitor data is shown in
Fig.~\ref{Monitor Raw Data and After ISP Data Result}, we find that
not only the color and contrast but also detail are very close to the
ground truth image. This proves that our pipeline can matches very
well with the smartphone ISP. The color and contrast are very natural
and very similar to the ground truth.  We can just replace the
degraded image with the restored image, and without change or adjust
too many other modules in the whole smartphone ISP.
 
As shown in Table~\ref{The Monitor Dataset Restoration Result}, after
processing by the ISP, the restored image even has a higher PSNR value
than the raw restored image. We think it is because some of the image
enhancement operations in the ISP, like the denoiser, perform further
improvements of the restored image. The CIEDE2000 is a metric that can
used to measure the color difference. We need to note that after ISP
processing, the CIEDE2000 metric is worse than for the raw
reconstruction. This illustrates further how sensitive the ISP can be
to even minute changes in the raw image.

\subsection{Real World Data Result}\label{Real world data}


\begin{table}[]
\centering

\resizebox{0.47\textwidth}{!}{%
\begin{tabular}{@{}lccccccccc@{}}
\toprule
               & \multicolumn{4}{c}{Monitor Pretrained Model} &                      & \multicolumn{4}{c}{Fine-tuning Model} \\ \cmidrule(lr){2-5} \cmidrule(l){7-10} 
               & PSNR     & SSIM    & LPIPS    & CIEDE2000    &                      & PSNR    & SSIM  & LPIPS  & CIEDE2000  \\ \midrule
Degraded Image & 28.42    & 0.86    & 0.140    & 0.056        &                      & 28.42   & 0.86  & 0.140  & 0.056      \\
PDCRN          & 26.70    & 0.71    & 0.224    & 4.261        & \multicolumn{1}{l}{} & 29.21   & 0.88  & 0.224  & 2.551      \\
Our            & 29.34    & 0.76    & 0.101    & 0.052        &                      & 31.40   & 0.91  & 0.128  & 0.039      \\ \bottomrule
\end{tabular}
}
\caption{Fine-tune before and after result}
\label{Fine-tune Before and After Result}
\end{table}

Moving on to real-world data, we first tested the monitor dataset
pretrained model on the real-world images without fine-tuning. The
result is shown in supplement Fig.~\ref{finetunelittleword} second column. As can
be seen, the small characters in the poster of the second column are a
little blurry -- even a little more blurry than the input image. All the papers based on ECCV 2020 UDC challenge ~\cite{sethumadhavan2020transform,sundar2020deep,ignatov2021learned,zhou2020image} are all only trained in the monitor-captured dataset, which we think may have similar problems. The reason may cause this and more detail about this is explained in the supplement. To solve the above imperfections of the monitor data,and further improve the result in the real practical application,  we fine-tune the training with real-world data. 
 
The fine-tuning result can be seen in the Supplement, Fig.~\ref{finetunelittleword}. The blur problem in the small characters is resolved,
which proves that our network after fine-tuning can address the
deblurring task.

\begin{figure}[h]
  \centering  
  \includegraphics[width=\linewidth]{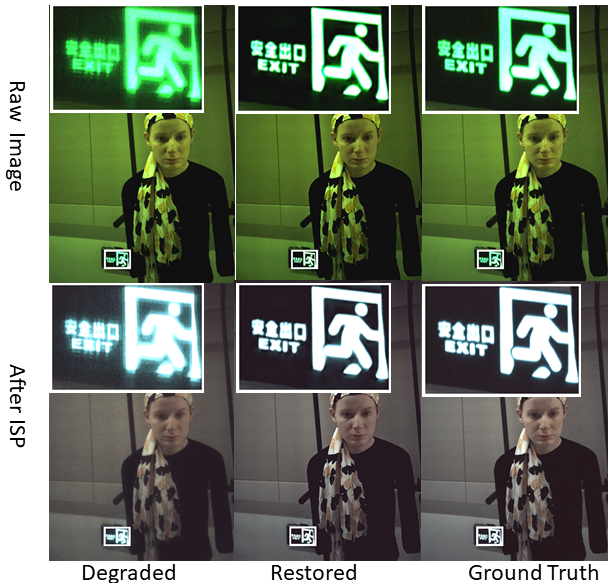}
  \caption{Real world raw data and ISP-processed result. The green
    image is the raw image. It is demosaicked raw data image. The
    ISP-proceed image is the image that go through ISP. Our restored
    image match with the Huawei ISP.}  
  \label{RealWorldDataRestorationResult}
\end{figure}

\begin{figure*}[h]
  \centering
  \includegraphics[width=\linewidth]{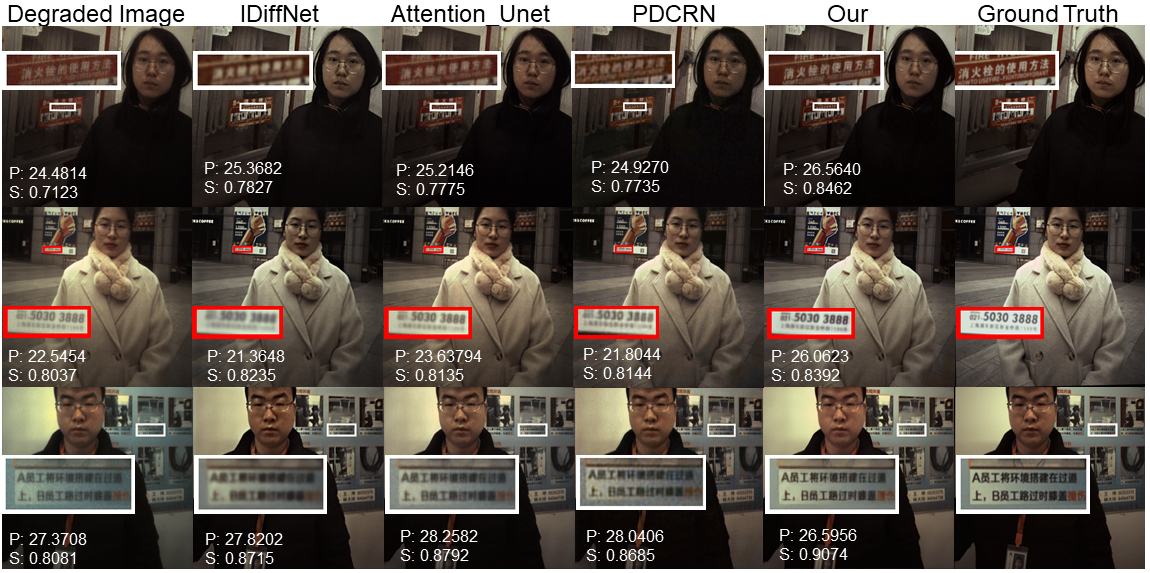}
  \caption{Quantitative comparison with state-of-art methods. We can
    see from the figure that our method restores more detail in the
    small characters. The PSNR and SSIM is computed in the ISP-proceed
    image.}
  \label{comparison}
\end{figure*}

After passing through our restoration pipeline and the smartphone ISP,
the final result is shown in the Fig.~\ref{RealWorldDataRestorationResult}. As
shown in the figure, some diffraction artifacts can be observed from
Fig.~\ref{RealWorldDataRestorationResult} "EXIT" logo in the degraded
image of the second example. There are some stripes in the bright green words. They are
removed in the restored image. The detail of the restored image is
almost the same as ground truth.  In addition, because the noise model
is related to the sensor, most of the time the denoising module would be
deployed in the raw domain. Therefore, in our raw domain algorithm, we
also consider the noise problem, and did the denoising work as
well.  We note
that the PSNR and SSIM results for the real world data are not quite
as good as for the monitor data, although the visual quality is very
good. We believe this may in part be due to residual misalignment in
the real world image pairs giving artificially lower ratings.

We did a comparison for some state-of-art methods similar to our
task. The IDiffNett~\cite{li2018imaging} and
AttentionUNet~\cite{oktay2018attention} are some state-of-art
dehazing or descattering neural residual network. Note that we use VGG
loss plus SSIM loss in IDiffNet and AttentionUNet. All have been pretrained and fine-tunes using our dataset.
We can see from
Fig.~\ref{comparison}, that our method restored more detail of the
small character. Table~\ref{comparison_table} is the comparison result
with preview dehazing or descattering neural network. We can find that
our method has the best PSNR and SSIM result. Note that
all metrics in Table~\ref{comparison_table} are computed based on the
raw image.

\begin{table}[]
\centering
\resizebox{0.85\width}{!}{%
\begin{tabular}{lccc}
\hline
               & PSNR $\uparrow$  & SSIM $\uparrow$ & LPIPS          $\downarrow$ \\ \hline
Degraded Image & 27.6735       & 0.8506       & 0.2024                    \\
IDiffNet       & 28.9234       & 0.8722       & 0.1737                    \\
AttentionUNet  & 29.1389       & 0.9018       & 0.1420                    \\
PDCRN          & 29.2126       & 0.8813       & 0.2241                    \\
Ours           & \textbf{31.2155} & \textbf{0.9023} & \textbf{0.0994}              \\ \hline
\end{tabular}
}
\caption{State-of-Art Dehazing Method Comparison. The parameter is the parameter of whole model. All metrics are computed by $3072 \times 2048$ size image.}
\label{comparison_table}
\end{table}

%

\section{Conclusion}
In this paper, we solve a real-world image-reconstruction problem and
support high image quality from under-display cameras, a new emerging
camera technology for consumer devices.  We propose a new raw-to-raw
pipeline that allows us to make use of the off-the-shelf smartphone
ISP without modifications. Instead of directly training on
ISP-processed image datasets, our raw-to-raw pipeline can avoid
situations where the degraded image and ground truth image go to
different branches of the ISP which would tremendously complicate the
training of the reconstruction network.

In addition to this core contribution, we also develop innovative
solutions for data collection using an HDR display, and for
synthesizing HDR portrait training data. 

Our pipeline directly replaces the degraded raw image with the
restored raw image without changing other modules. Our pipeline has
very good performance on the raw data image, an crucially it still has
good performance after applying the ISP to the restored image. The
result has a very natural color, contrast, and detailed
information. This proves that our pipeline can cooperate with the
current off-the-shelf ISP. We believe that this raw-to-raw paradigm
can be useful for many other image restoration tasks con consumer
devices in the future.

{\small
\bibliographystyle{ieee_fullname}
\bibliography{egbib}
}

\pagebreak

\textbf{ISP-Agnostic Image Reconstruction for Under-Display Cameras Supplement}

\beginsupplement

\begin{figure}[h!]
  \centering
  \includegraphics[width=\linewidth]{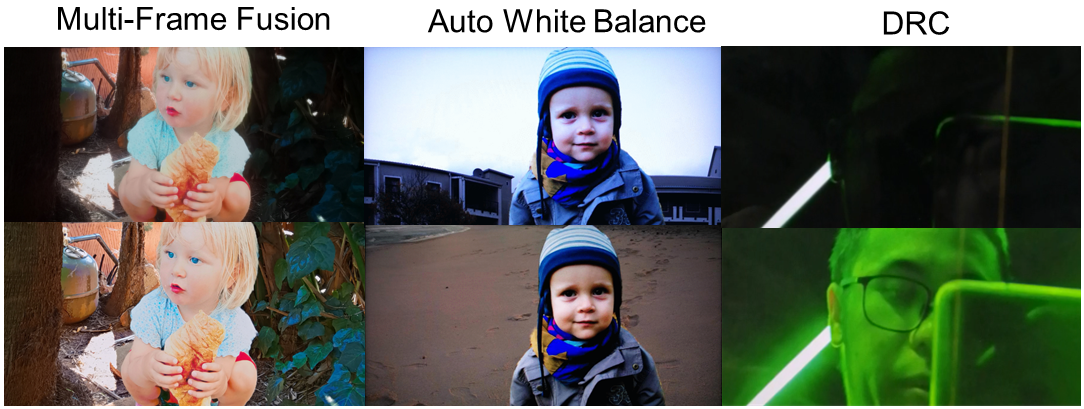}
  \caption{ISP non-linearity. The first column is brightness change caused by smartphone ISP multi-frame fusion. The dark part of background become bright after multi-frame fusion. The second column is smartphone ISP auto white balance operation. The boy face change from blue to yellow. The third column is smartphone ISP dynamic range correction, the image change from dark to bright.}  
  \label{The ISP nonlinear operation}
\end{figure}

\section{ISP Nonlinearities}
In Fig.~\ref{The ISP nonlinear operation} in the paper, we show several examples of
how the ISP processing can affect the final appearance of the image.
The first column shows an image pair captured from an HDR monitor. The
bottom image was captured with multi-frame fusion, while the top image
was shown without this operation.  As can be seen, the leaf and tree
trunk on the top is dark. However, on the bottom one, the same objects
appear brighter. The change happens specifically because of the
process of multi-frame fusion. Every pair - dark and bright area of
the image – is different and the brightness change cannot be predicted
without a detailed model of the ISP algorithms.

The second column depicts images that were captured of boy in front of
different backgrounds. The portrait of the boy was composited over
different backgrounds, and again shown on an HDR monitor for capturing
(see Sec.~\ref{Data Augmentation}). We note that the ISP reproduces
different facial colors for the two images due to auto white
balancing. Color distortions also exist between UDC and regular image
pairs, so this behavior is again problematic for a training dataset.

Finally, the third column is example a high dynamic range scene. In
this case, the image undergoes a dynamic range correction (DRC). This
will also break the brightness consistency between the degraded image
and the reference image.

\begin{figure}[h]
  \centering
  \includegraphics[width=\linewidth]{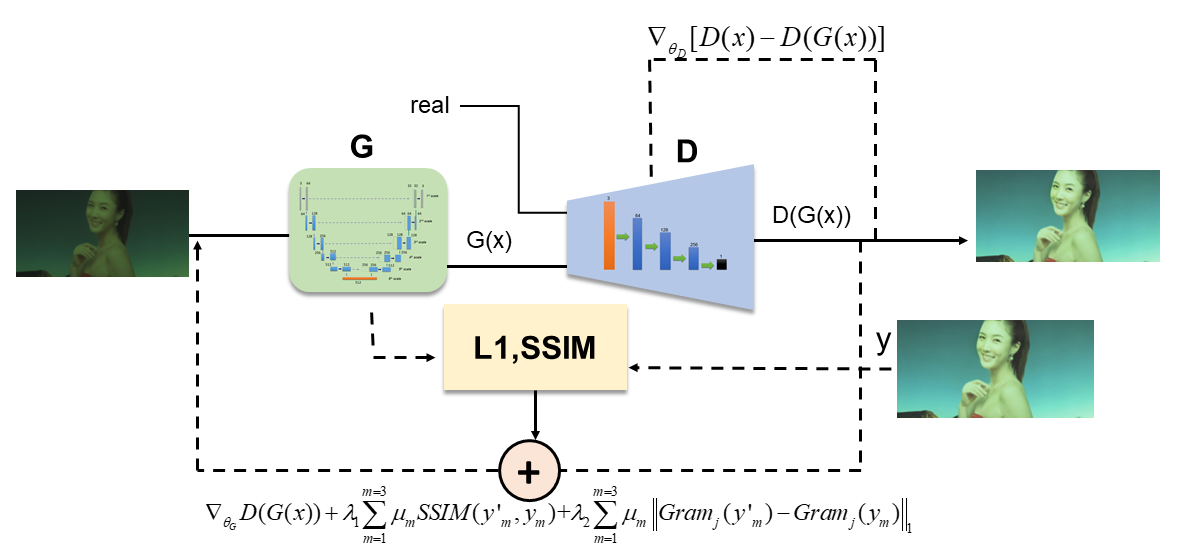}
  \caption{WGAN-GP Structure. $G$ and $D$ respectively stand for the
    generator and the discriminator. We use an optimized UNet
    structure for the generator, and a simple CNN for the
    discriminator. We add a content loss composed of SSIM and
    perceptual loss to constrain the generator.\label{fig:network}}
\end{figure}

\begin{figure}[h]
  \centering
  \includegraphics[width=\linewidth]{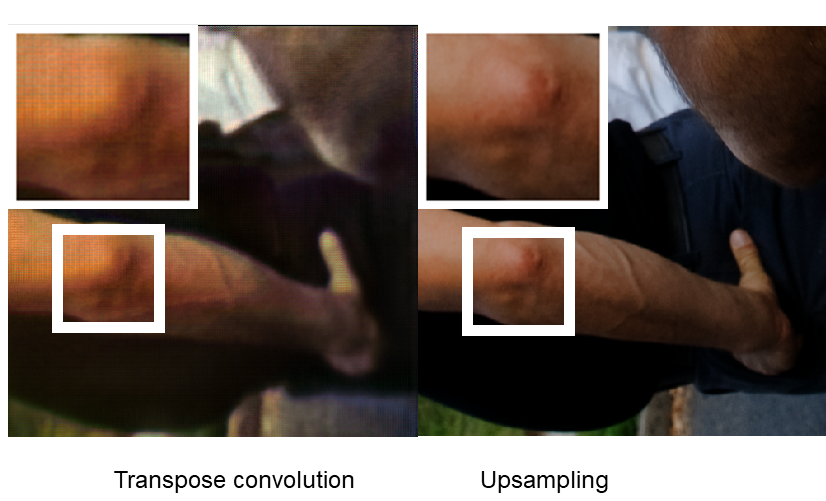}
  \caption{Transpose convolution block artifacts. The left image is
    restored result of transpose convolution layer. We can see clearly
    that there is a block artifacts in the left image. On the right is
    the restored result with the upsampling layer. It is much more
    clean and shows no block artifacts. }
  \label{checkernoise}
\end{figure}

\section{Dataset}
\begin{figure}[h]
  \centering
  \includegraphics[width=\linewidth]{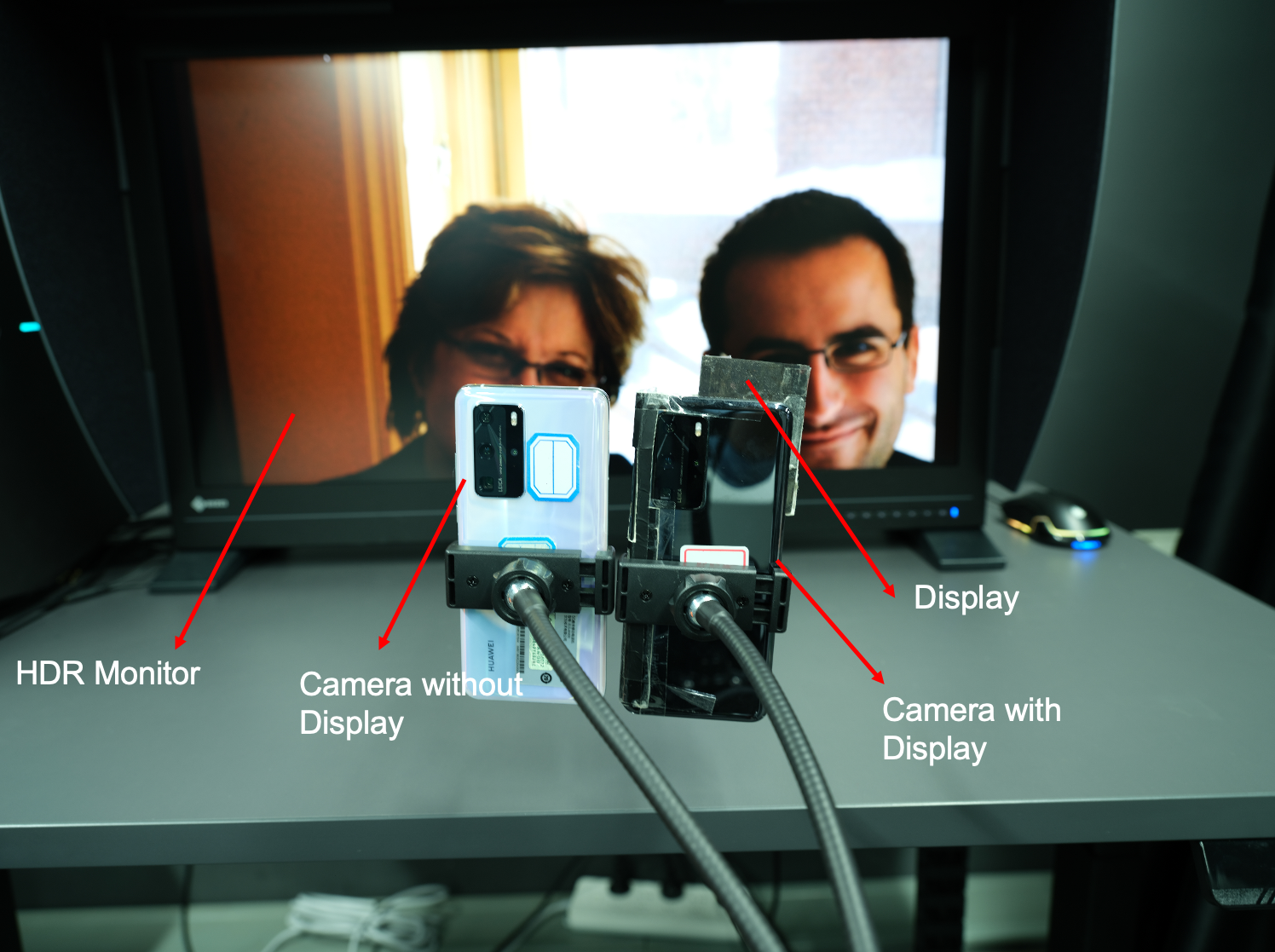}
  \caption{HDR Monitor data collection setup. Two mobile phones are
    placed in front of the HDR monitor, one with the UDC hardware, and
    one with an identical reference camera without the UDC feature.
    The phone is oriented vertically to mimic the most common usage
    case. The setup is in a dark room, and we fix the exposure time
    and iso setting of the cameras.}
  \label{Monitor Data Collection Setup}
\end{figure}

\subsection{Calibration}
Before capturing a large number of image pairs for training, we
calibrated the system as follows: (1) we displayed checkerboard
patterns on the monitor and then estimated the camera lens's
distortion parameters and the homography transform matrix between each
camera and the monitor. (2) we captured the white image displayed on
the monitor to compensate for the brightness non-uniformity effect of
the monitor and the two cameras' vignetting effect.  (3) We calibrated
the display color of the HDR monitor. Since ordinary chromatic meter
is unable to measure HDR signal, we adopted a complementary
measurement method by using a DSLR Canon Camera 5D MarkIV and a
spectral camera Specim IQ. We utilized the Canon camera to capture a
stack of color sample images shown on the monitor with varying
exposure time, and fused them to obtain the HDR color responses of the
camera. We then used the spectral camera to measure the spectral
distribution functions of the primaries of the monitor, and calculate
a color transform matrix from the color responses of the Canon camera
to CIE RGB values, and obtain the 3x3 color transform matrix from CIE XYZ color space to the Canon camera by capturing the three same images, and finally obtain the xyz value from the arbitrary captured color. After capturing and monitor image, we did image registration and apply this color transform to map the captured image to correct HDR color.

\begin{figure}[htp]
    \centering
    \includegraphics[width=\columnwidth]{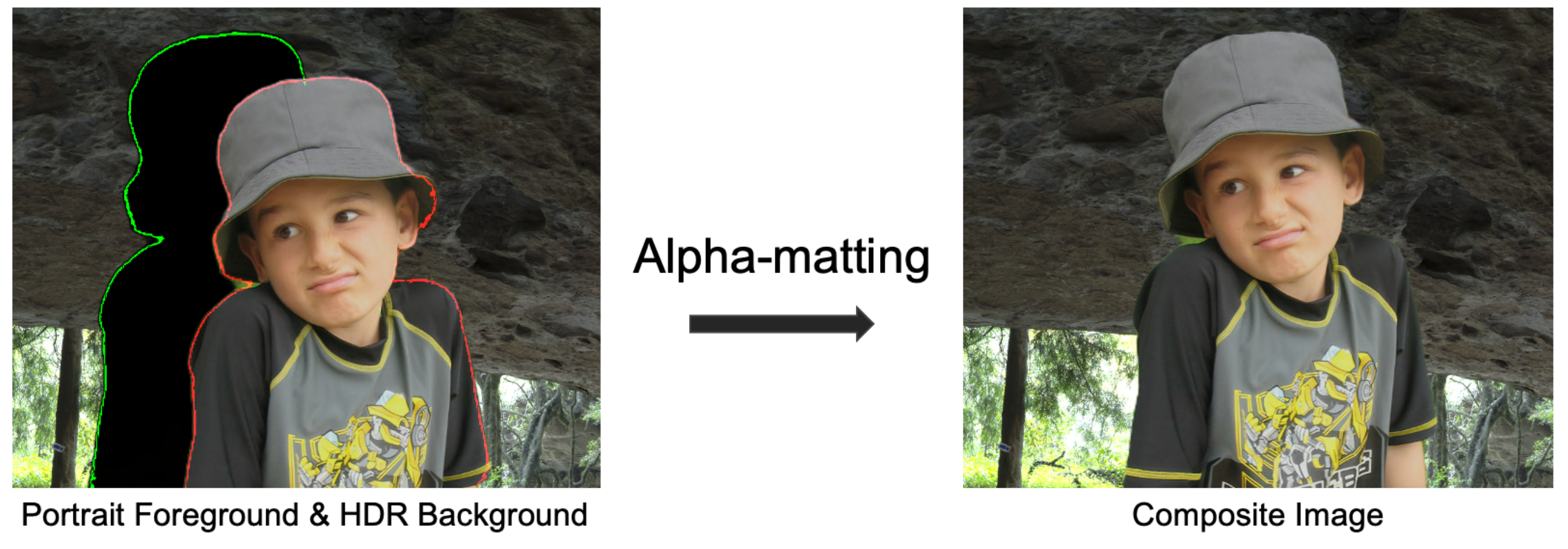}
    \caption{Portrait Image Composition.}
   \label{fig:composite}
\end{figure}

\subsection{HDR Portrait Dataset Upsample}\label{HDR Portrait Dataset Upsample}
\begin{figure*}[h]
  \centering
  \includegraphics[width=\linewidth]{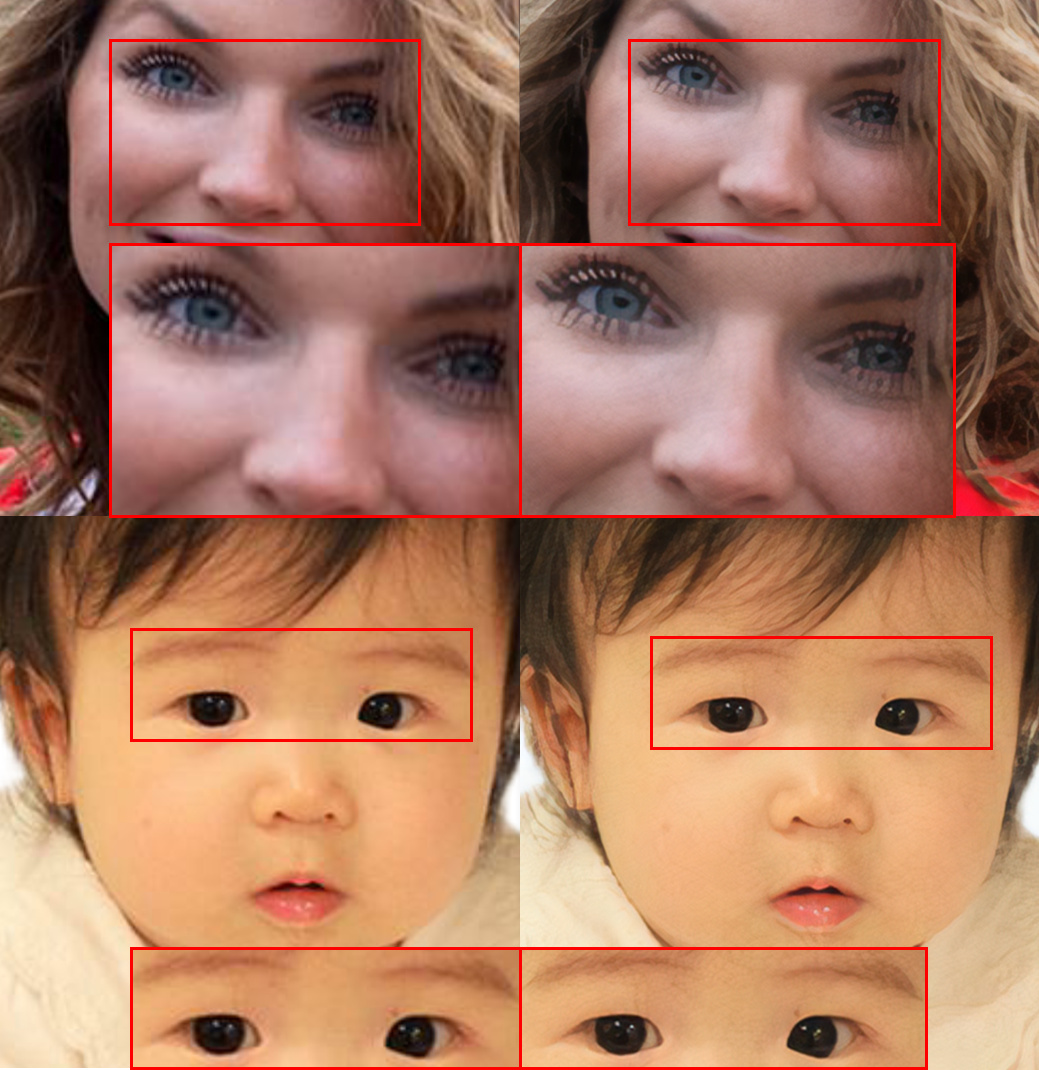}
\caption{Upsample}
  \label{Portrait Image Upsample Comparison}
\end{figure*}

To fill the gaps
between the resolution of the portrait images and HDR images, we
utilized a portrait superresolution method~\cite{chen2018fsrnet} to
enhance the resolution of portrait images and match the resolutions of
the portrait and background. The result shown in the Fig. \ref{Portrait Image Upsample Comparison}
The network we used for superresolution enhance the resolution with an upsampling factor of 4. It will introduce some artifacts but these artifacts is the same for the UDC image and no UDC image, therefore, it won't cause any problem for our restoration task.
The foreground and background of the
chosen images are harmonious and hard to be distinguished for human
eyes. The images include indoors and outdoor scenes under bright and
dark light condition.

\subsection{Brightness Normalization and Compensation} \label{Brightness Normalization and Compensation}

Because the display absorbs some light, the shutter speed for camera
with display is longer than the camera without display. Moreover, we
found it difficult to disable the autoexposure on the mobile phone
prototypes we used. Instead, we used image metadata to compensate for
any differences in exposure and iso values. Specifically, we use a
brightness compensation factor
\begin{equation}
\eta=\frac{t'}{iso}\frac{iso'}{t} \kappa,
\end{equation}
where $t$ is the exposure time of the image captured without display,
and $t'$ is the exposure time of the image captured with the
display. $iso$ is the ISO value of image captured without display, and
$iso'$ is the ISO value of the image captured with the display.
$\kappa$ is a global brightness compensation factor that approximately
compensates for the brightness loss in the UDC hardware, in our case
$\kappa=0.5$. This value is determined by tuning manually and find the best result that make two images brightness match with each other.

We divided this normalization factor $\eta$ for every degraded
image pixel value. Then we keep a relative fixed mapping relationship between the
input image and the output image, so that the neural network can learn
it easily.

Note that all the smartphone raw images have a black level
current. Before doing the brightness normalization and compensation we
subtract the black level current first. Then do the brightness
normalization and compensation based in the formula in the paper. Then
go through our restoration algorithm and convert to raw image
again. Finally we add the black level current back to match regular
RAW images and facilitate post-processing with the  smartphone ISP.

\subsection{Training and Test Detail}


All the neural networks are implemented in Pytorch 1.5 and Python
3.7. All the training is done on an Nvidia Tesla V100 GPU. Due to GPU
memory limitations, testing is performed on the CPU, since this allows
us to load a full sized image. The generator has 2.15M parameters, and
requires 160.30 GFLOPs to reconstruct a $3072 \times 2048$ image. The run
time is 0.25 s for a $3072\times2048$ image on the Intel i7-9750H
2.6GHz CPU with 32GB memory.

\section{Result}

\subsection{Ablation Study}

\begin{figure}[h]
  \centering
  \includegraphics[width=\linewidth]{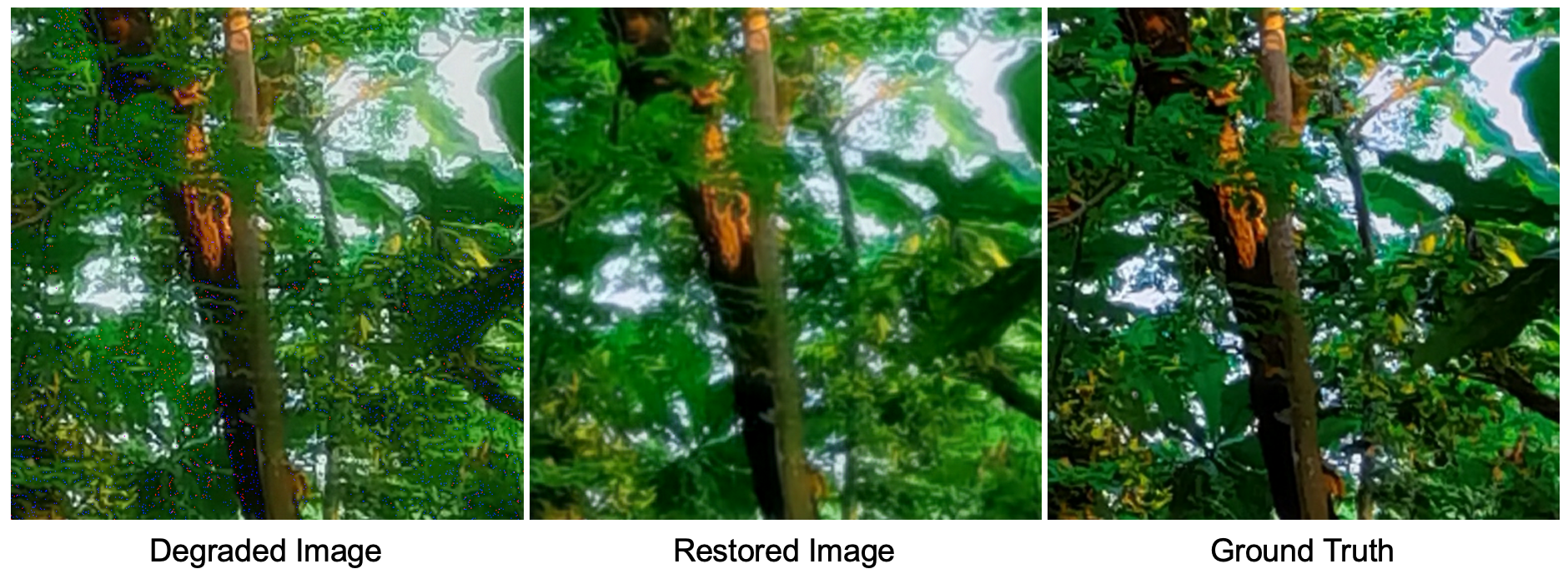}
  \caption{Training the UDC restoration network on ISP-processed
    image pairs does not produce good results due to the black-box
    non-linearities in the ISP. }
  \label{Direct Training Result of JPG Image}
\end{figure}

\begin{figure}[h]
  \centering
  \includegraphics[width=\linewidth]{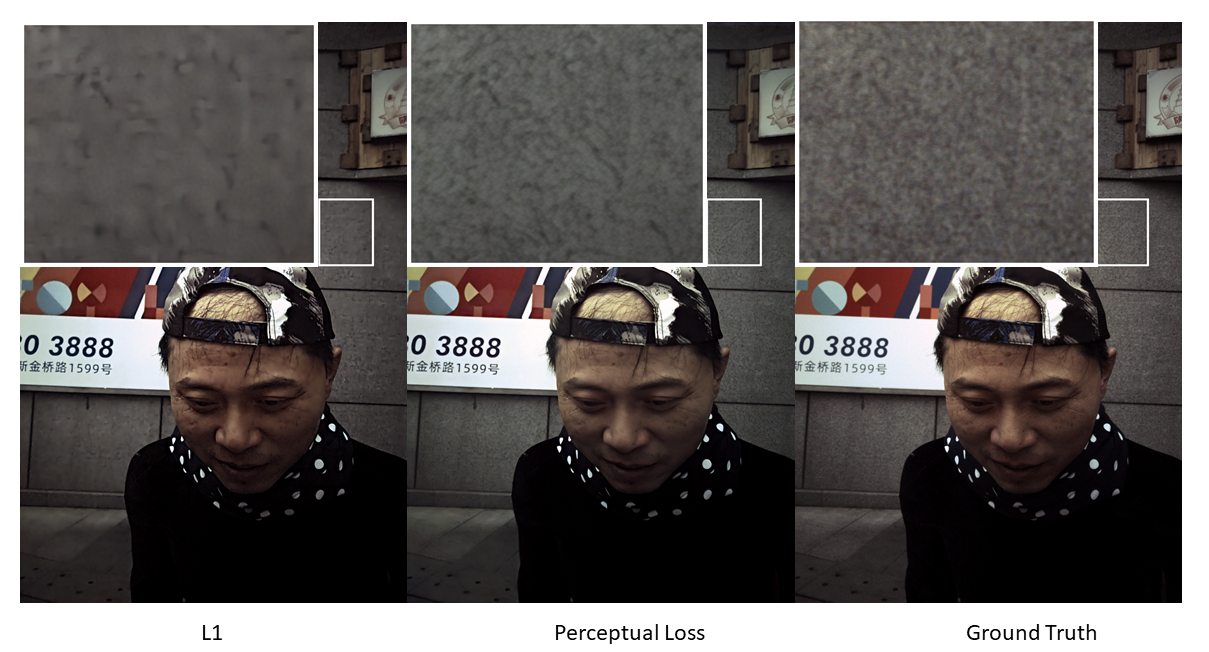}
  \caption{L1 Loss and perceptual loss comparison.}  
  \label{L1vsVGG}
\end{figure}

The reason that we choose perceptual loss is because perceptual loss will preserve more texture detail
and have better perceptual performance. As shown in
Fig.~\ref{L1vsVGG}, the perceptual loss restores more texture detail
than the L1 loss, especially in the wall region. In addition, the L1
loss is more sensitive to misalignment in the image pairs. As shown in
Table~\ref{Ablation Study}, the perceptual loss and SSIM loss
combination is better than the L1 and SSIM loss combination.

\begin{figure}[h!]
  \centering
  \includegraphics[width=\linewidth]{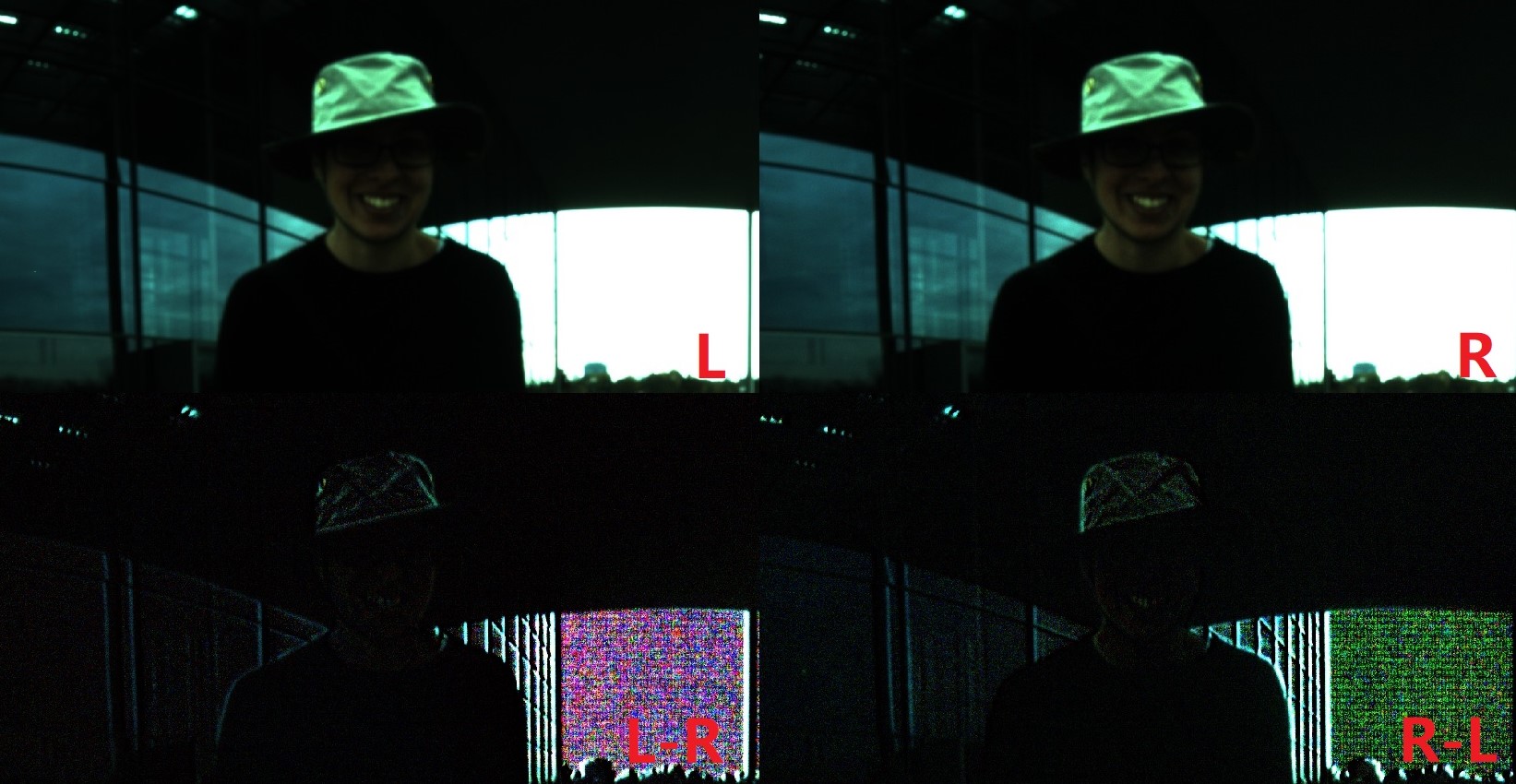}
  \caption{Monitor structured noise. This monitor image is captured by
    Cannon Mark IV dual pixel camera. $L$ and $R$
    respectively refer to the left and right subimages of the dual
    pixel sensor, and $L-R$ refers to the amplified image difference.}
  \label{structure noise}
\end{figure}

\begin{figure*}[h]
  \centering
  \includegraphics[width=\linewidth]{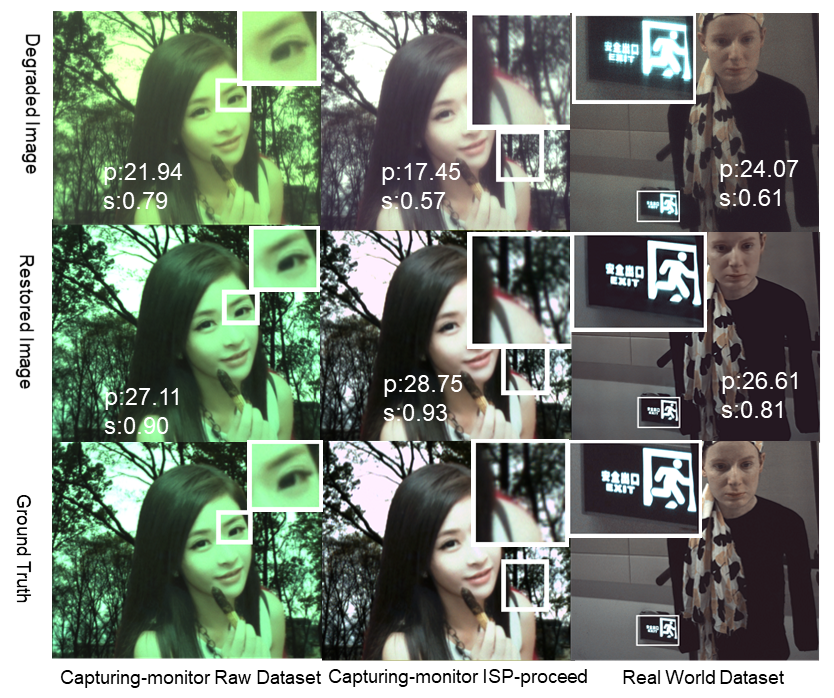}
\caption{Restoration result. The first column is the monitor raw
    dataset restoration result. The restored raw image is processed by
    the legacy ISP, and the output is shown in the second column. The
    third column is real-world test result. We can see that there is
    obvious stripe-like diffraction artifacts in the green words in
    the third column degraded image, but removed in our result.  }
  \label{threedatasetresult}
\end{figure*}

\subsection{Pretrain Dataset and Fine-tune}

\begin{figure*}[h!]
  \centering
  \includegraphics[width=\linewidth]{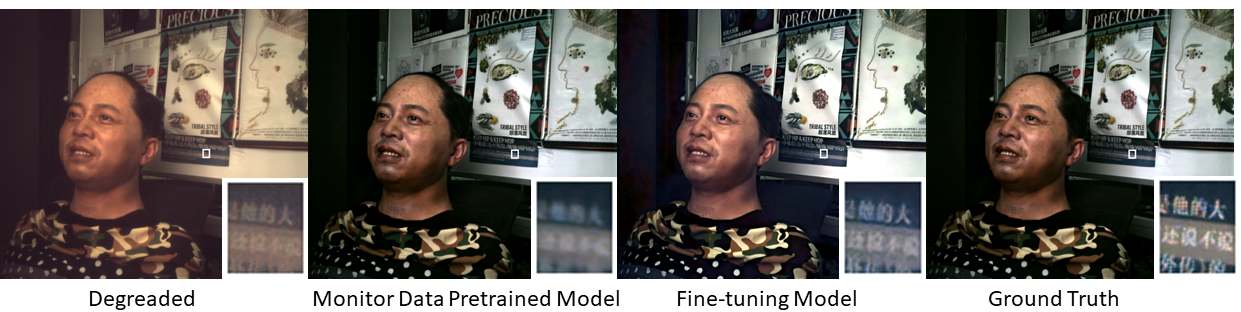}
  \caption{Effect of fine-tuning with real-world data. From left to
    right: input image, restoration result for monitor pre-trained
    model, and restoration result using the full fine-tuned model. }  
  \label{finetunelittleword}
\end{figure*}

We want to note that the zoom in patch in the Fig. \ref{finetunelittleword} is very small in the full image. The monitor pretrained model performance is still overall very good. However, there is a very tiny blurry for the little detail. The Table \ref{Fine-tune Before and After Result} shows the result of monitor pretrained model before and after fine-tune. The PDCRN is the method ~\cite{sethumadhavan2020transform} in the UDC challenge. We can see that both our method and PDCRN has the same problem that although monitor pretrained model can perform very well in the monitor dataset, the metrics number decrease a lot when it test in the real world dataset.  There
are several reasons that may account for this. First of all, the
monitor image resolution is limited. It can not be as sharp and
detailed as the real-world image, and therefore small texture details
are lost. Second, the monitor image suffers from some structured
noise. The Fig.~\ref{structure
  noise} is a captured image of HDR monitor used for creating the monitor dataset. The top row are images captured by dual pixel sensor. The bottom row are the images that difference between left view image and right view image.  Consider the white sky in the background of Fig.~\ref{structure
  noise}. This area should be very smooth with almost constant pixel
values. However the difference of two views shows there is structured noise for monitor.

\section{ISP Post-processing}

For ISP processign after reconstruction, we utilize two smartphone
ISPs. First is the black-box Huawei smartphone ISP, while another is
based on a simple ISP pipeline built by ourselves. In the latter The
demosaicing is based on the gradient-corrected linear
interpolation. The auto white balance correction algorithm is based on
the paper~\cite{Afifi2019WBsRGBImages}. Finally, we apply gamma
correction. Our reconstructions perform equally well under both of
these ISPs.

%
%
%
%
%
%
%
%
%
%

\begin{figure*}[h]
  \centering
    \includegraphics[width=\linewidth]{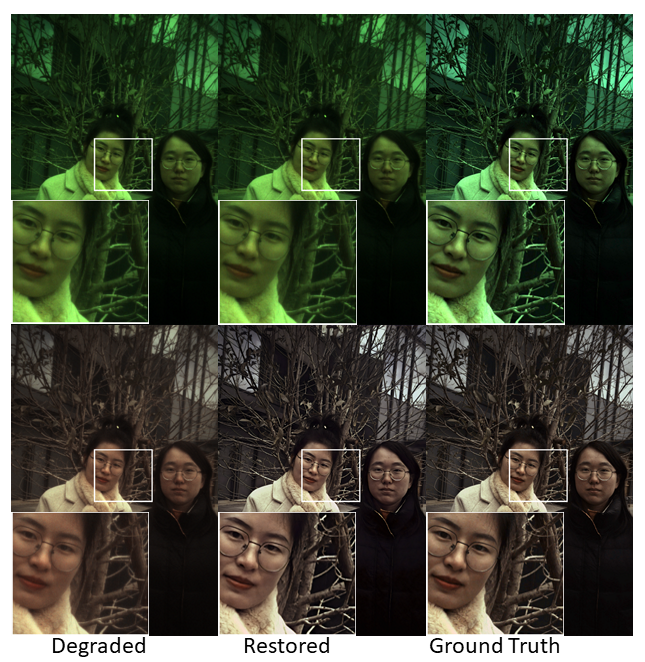}
   \caption{Some More Real World Data Results.} 

\end{figure*}

\begin{figure*}[h]
  \centering
    \includegraphics[width=\linewidth]{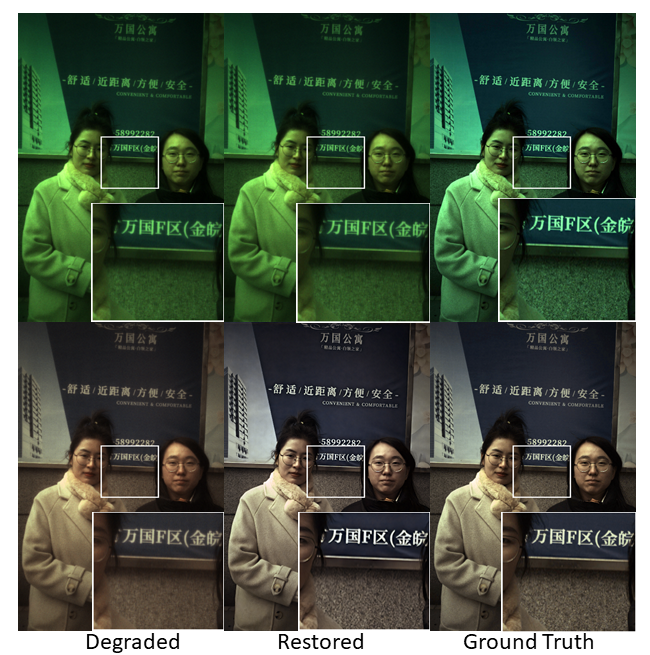}
   \caption{Some More Real World Data Results.} 

\end{figure*}

\begin{figure*}[h]
  \centering
    \includegraphics[width=\linewidth]{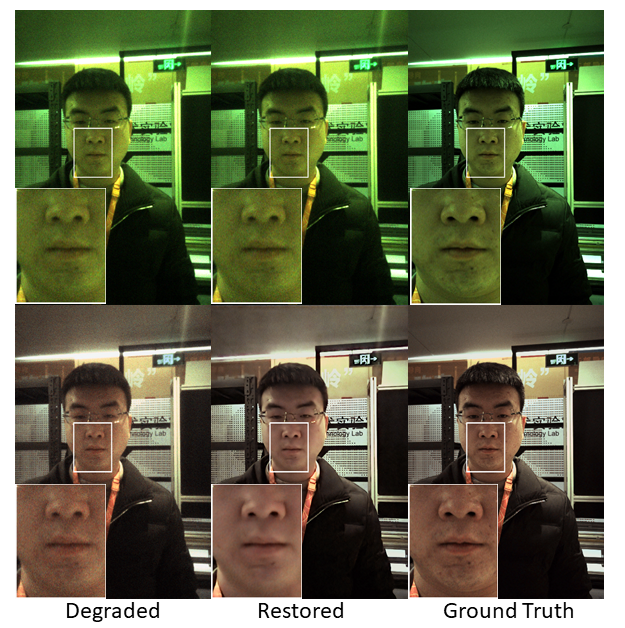}
   \caption{Some More Real World Data Results.} 

\end{figure*}

\begin{figure*}[h]
  \centering
    \includegraphics[width=\linewidth]{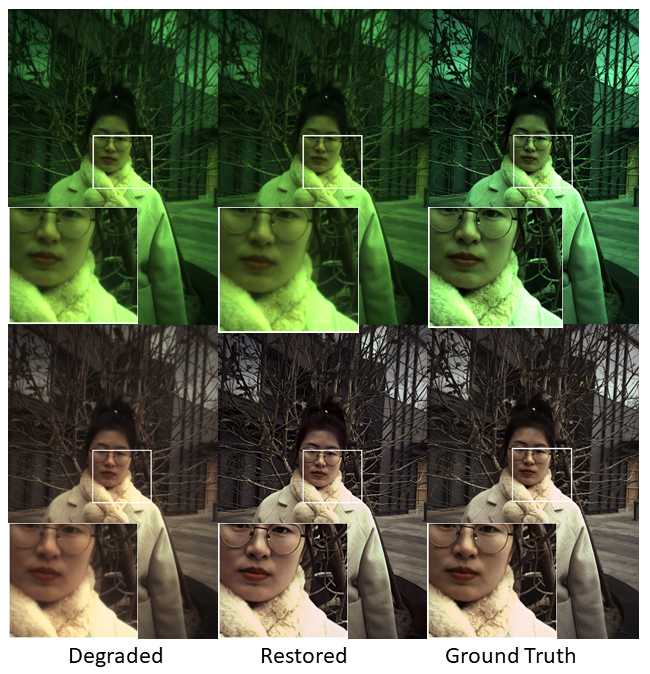}
   \caption{Some More Real World Data Results.} 

\end{figure*}

\begin{figure*}[h]
  \centering
    \includegraphics[width=\linewidth]{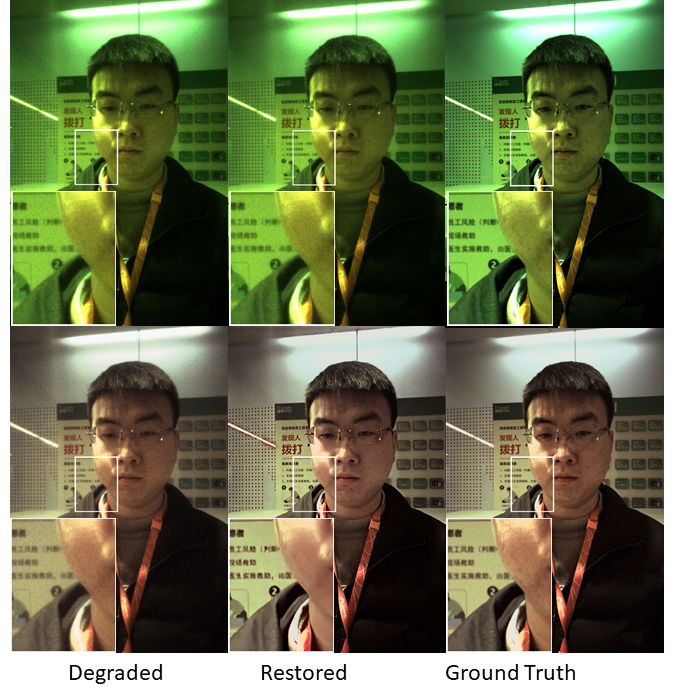}
   \caption{Some More Real World Data Results.} 

\end{figure*}

\begin{figure*}[h]
  \centering
    \includegraphics[width=\linewidth]{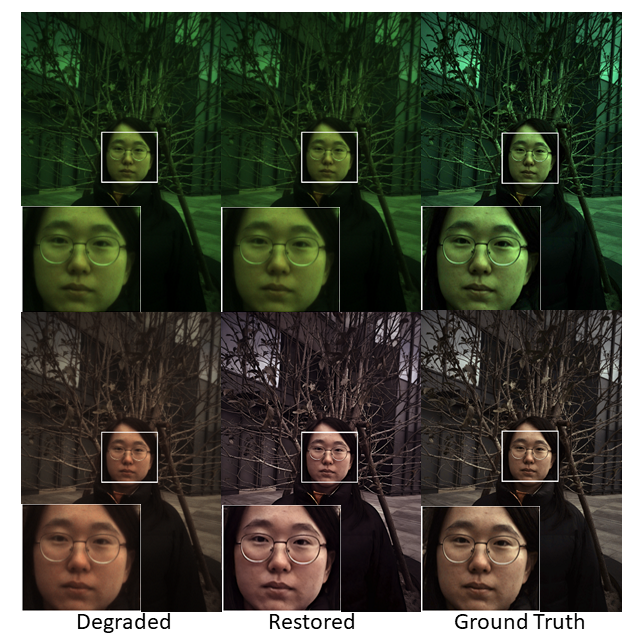}
   \caption{Some More Real World Data Results.} 

\end{figure*}

\begin{figure*}[h]
  \centering
    \includegraphics[width=\linewidth]{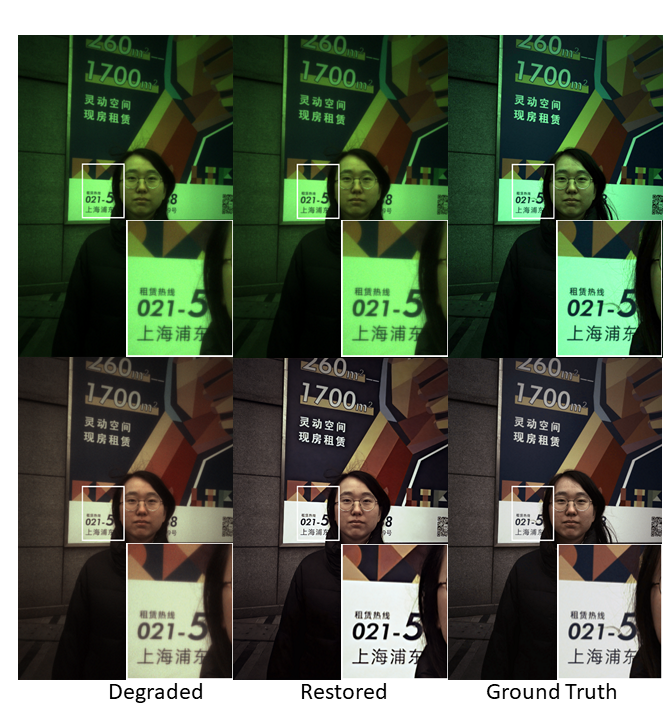}
   \caption{Some More Real World Data Results.} 

\end{figure*}

\begin{figure*}[h]
  \centering
    \includegraphics[width=\linewidth]{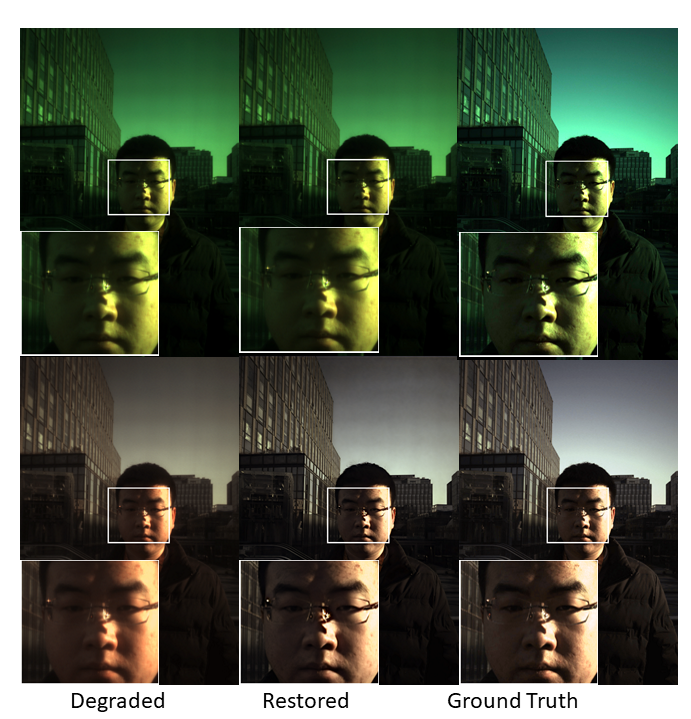}
   \caption{Some More Real World Data Results.} 

\end{figure*}

\begin{figure*}[h]
  \centering
    \includegraphics[width=\linewidth]{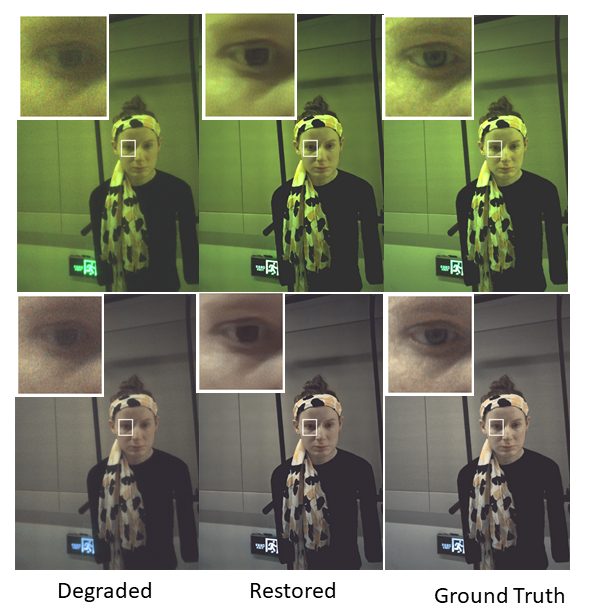}
   \caption{Some More Real World Data Results.} 

\end{figure*}

\begin{figure*}[h]
  \centering
    \includegraphics[width=\linewidth]{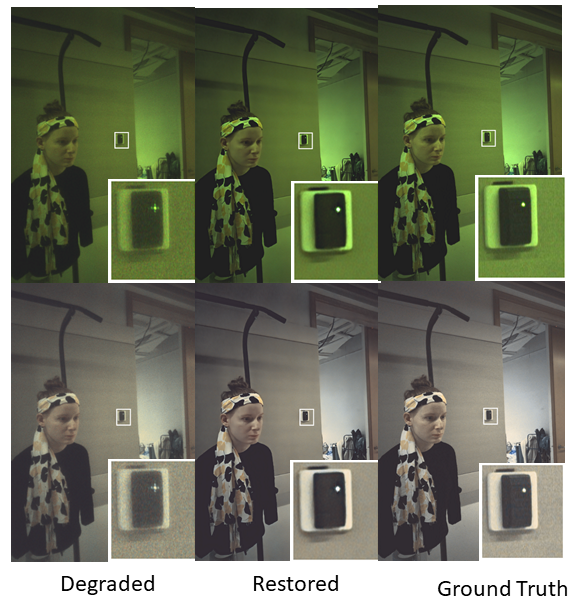}
   \caption{Some More Real World Data Results.} 

\end{figure*}

\begin{figure*}[h]
  \centering
    \includegraphics[width=\linewidth]{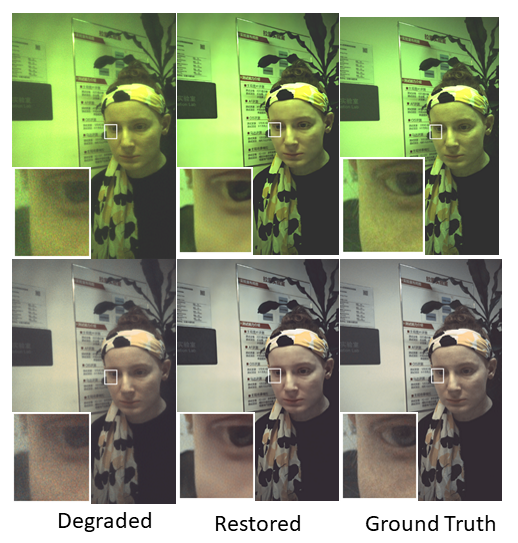}
   \caption{Some More Real World Data Results.} 

\end{figure*}

\begin{figure*}[h]
  \centering
    \includegraphics[width=\linewidth]{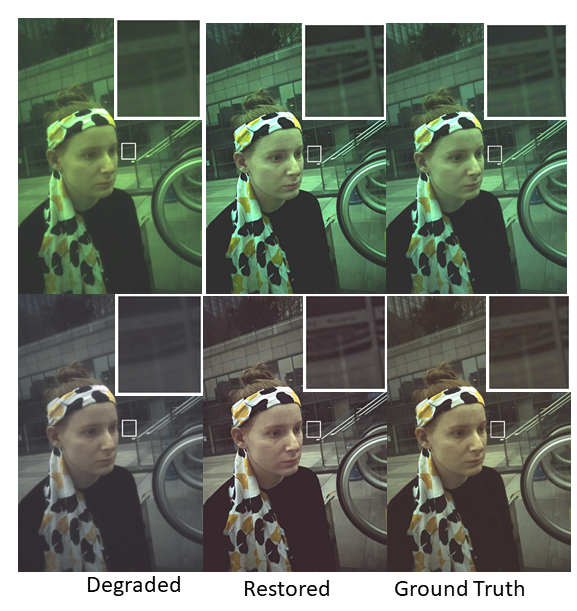}
   \caption{Some More Real World Data Results.} 

\end{figure*}

\begin{figure*}[h]
  \centering
    \includegraphics[width=\linewidth]{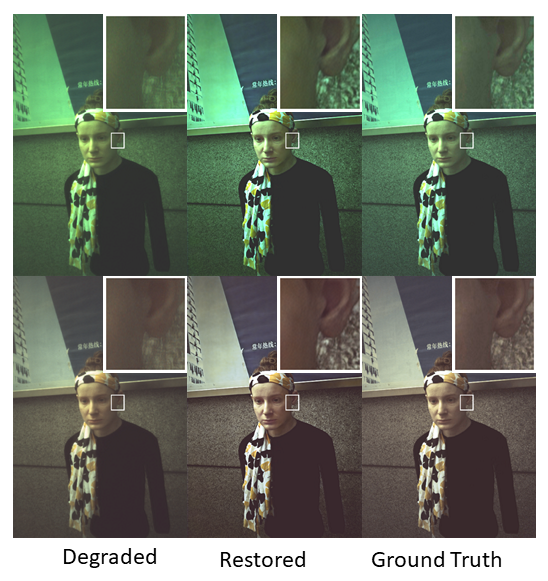}
   \caption{Some More Real World Data Results.} 

\end{figure*}

\begin{figure*}[h]
  \centering
    \includegraphics[width=\linewidth]{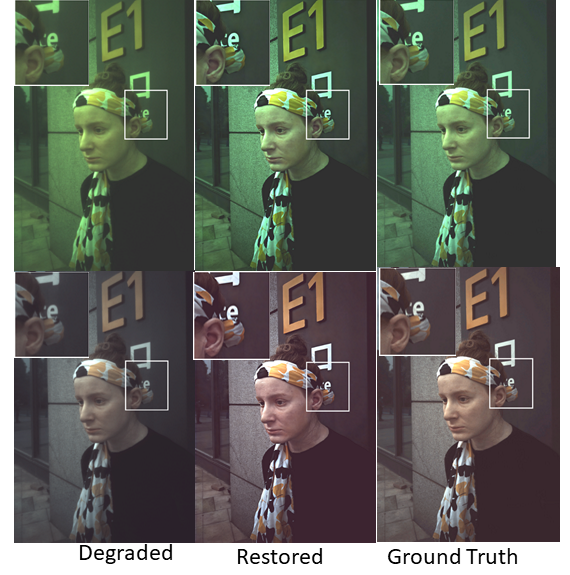}
   \caption{Some More Real World Data Results.} 

\end{figure*}

\begin{figure*}[h]
  \centering
    \includegraphics[width=\linewidth]{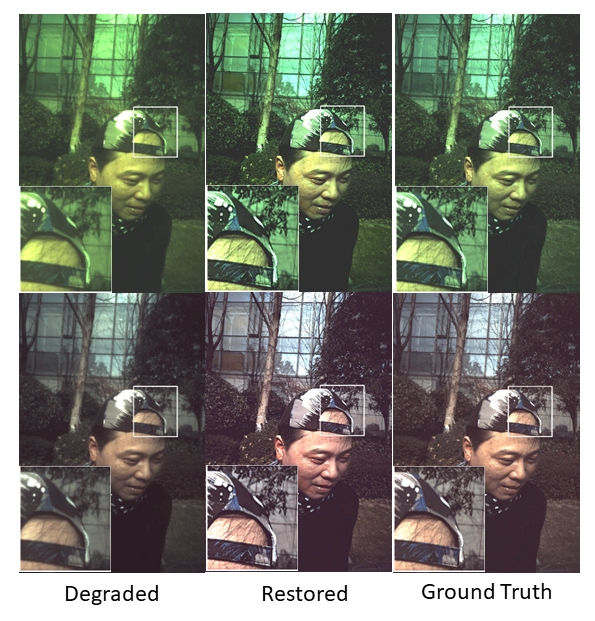}
   \caption{Some More Real World Data Results.} 

\end{figure*}

\begin{figure*}[h]
  \centering
    \includegraphics[width=\linewidth]{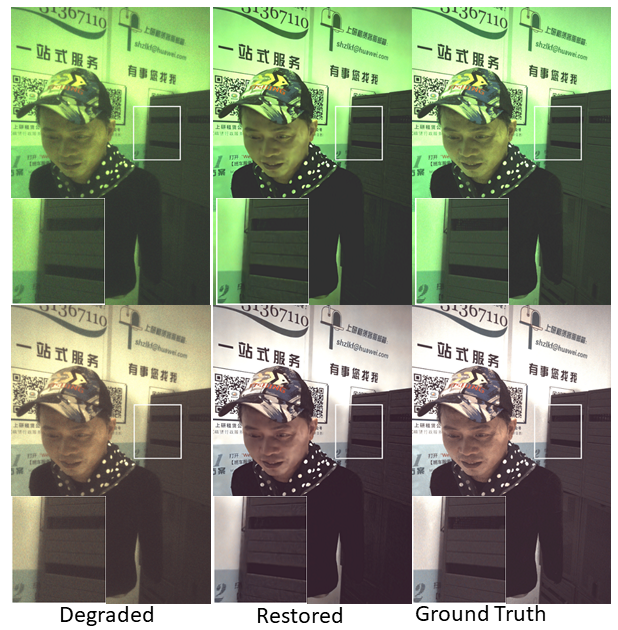}
   \caption{Some More Real World Data Results.} 

\end{figure*}

\begin{figure*}[h]
  \centering
    \includegraphics[width=\linewidth]{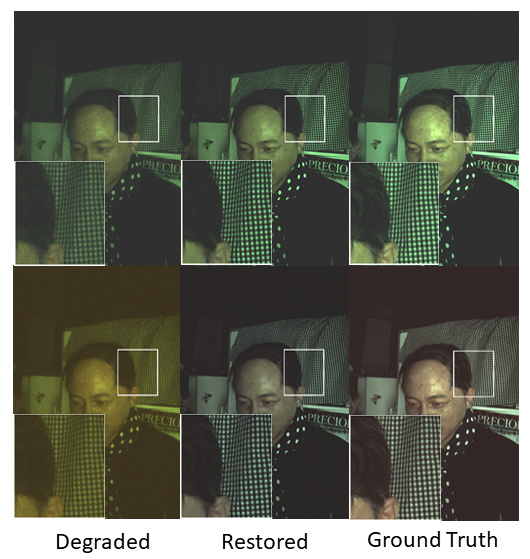}
   \caption{Some More Real World Data Results.} 

\end{figure*}

\begin{figure*}[h]
  \centering
    \includegraphics[width=\linewidth]{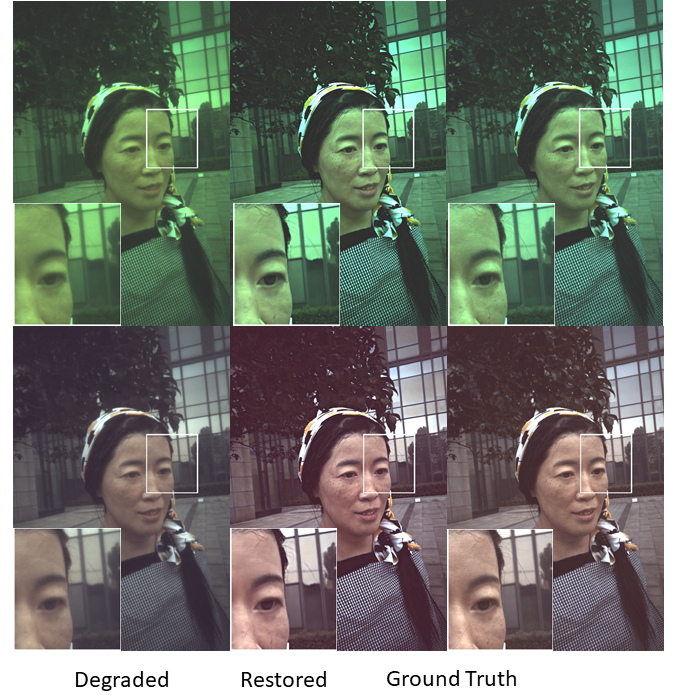}
   \caption{Some More Real World Data Results.} 

\end{figure*}

\begin{figure*}[h]
  \centering
    \includegraphics[width=\linewidth]{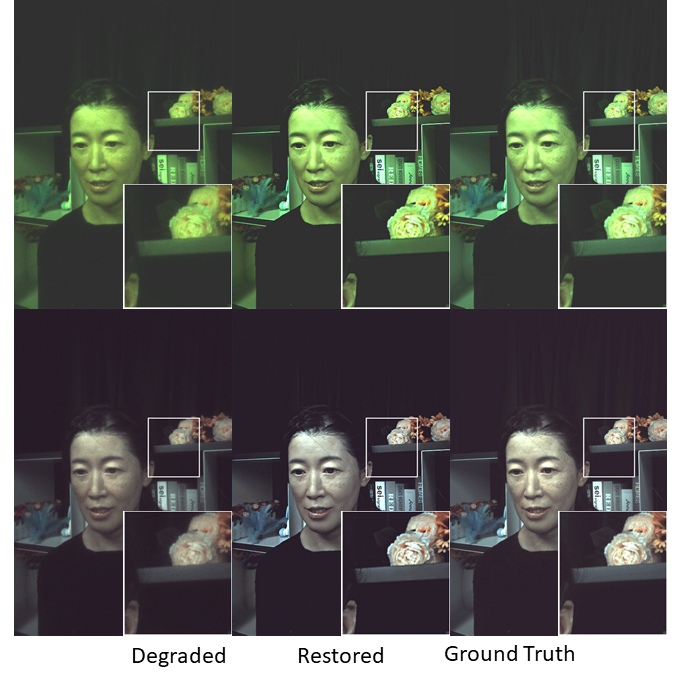}
   \caption{Some More Real World Data Results.} 

\end{figure*}

\begin{figure*}[h]
  \centering
    \includegraphics[width=\linewidth]{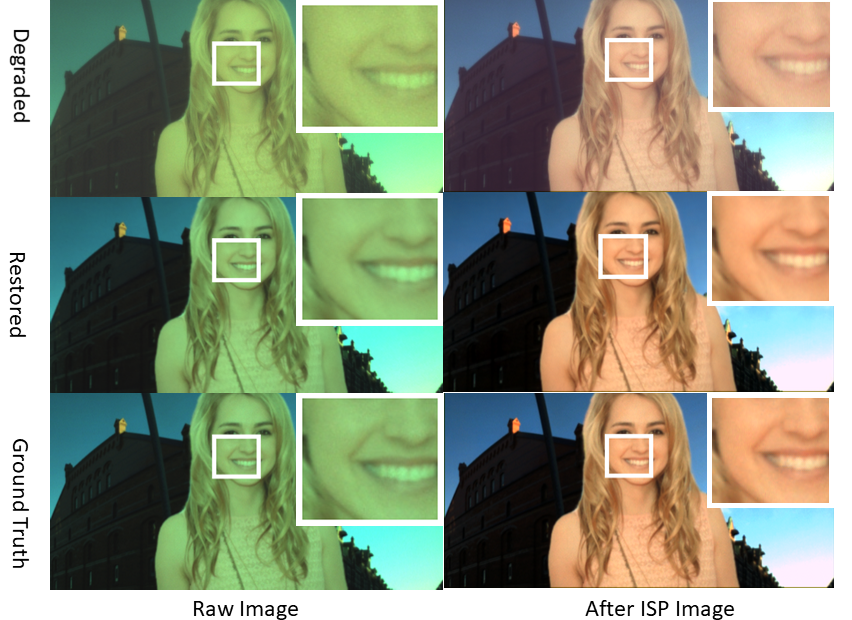}
   \caption{Some More Monitor Data Results.} 

\end{figure*}

\begin{figure*}[h]
  \centering
    \includegraphics[width=\linewidth]{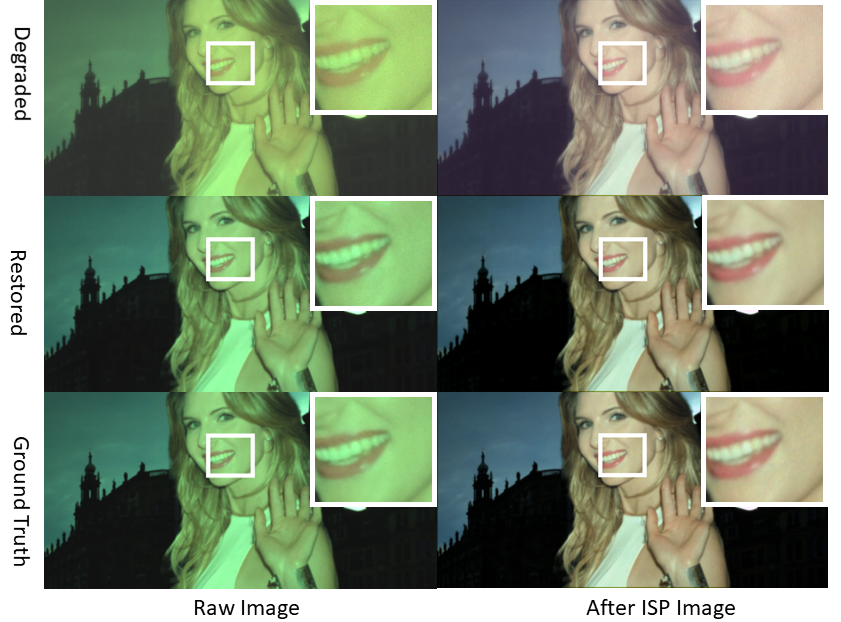}
   \caption{Some More Monitor Data Results.} 

\end{figure*}

\begin{figure*}[h]
  \centering
    \includegraphics[width=\linewidth]{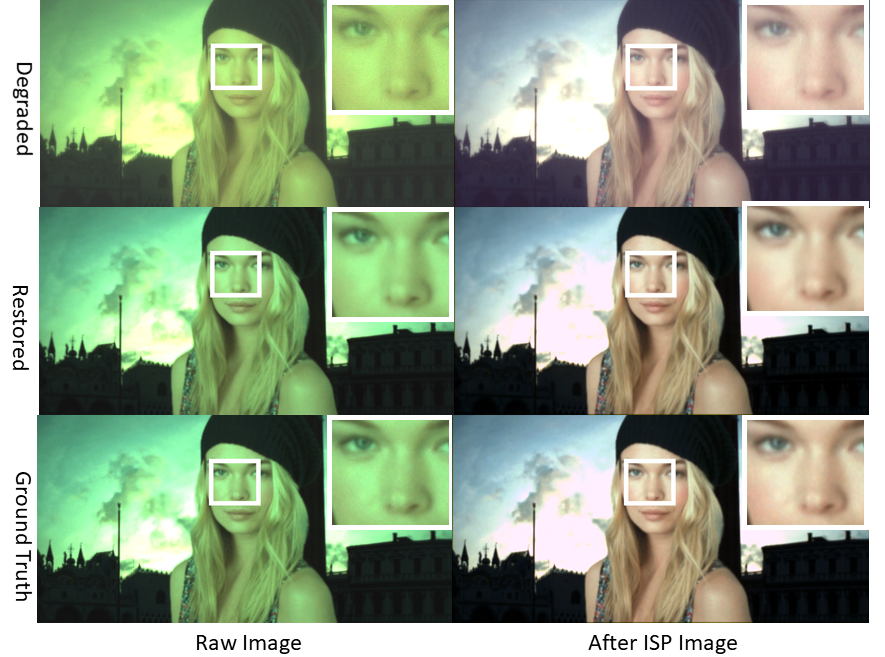}
   \caption{Some More Monitor Data Results.} 

\end{figure*}

\begin{figure*}[h]
  \centering
    \includegraphics[width=\linewidth]{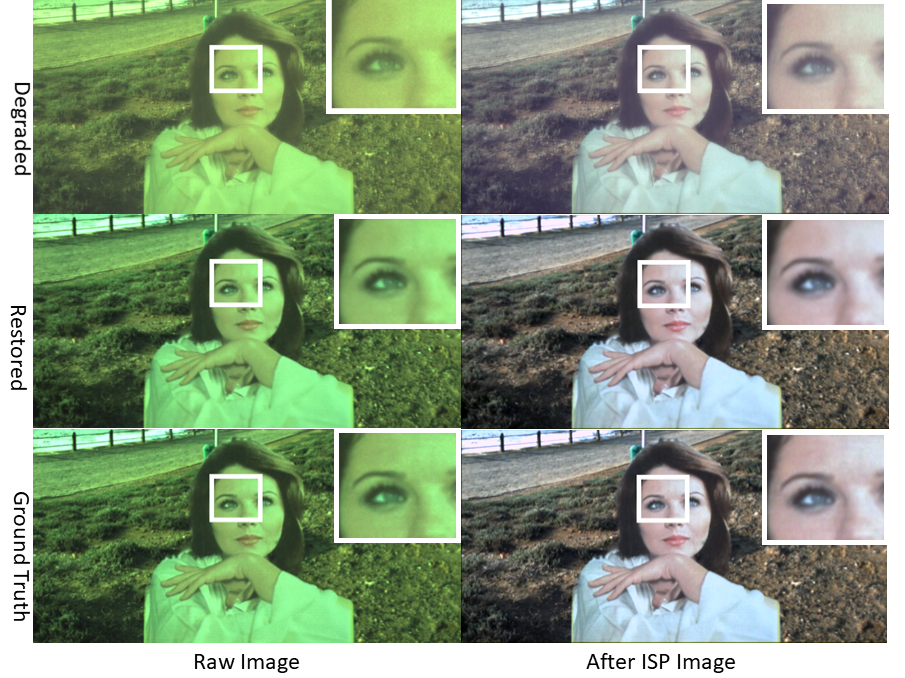}
   \caption{Some More Monitor Data Results.} 

\end{figure*}

\begin{figure*}[h]
  \centering
    \includegraphics[width=\linewidth]{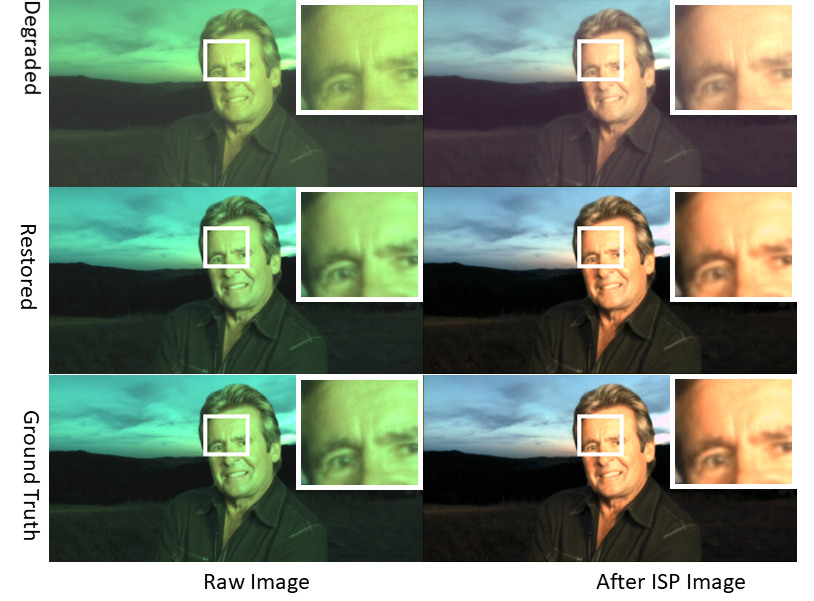}
   \caption{Some More Monitor Data Results.} 

\end{figure*}

%
%


\end{document}